\documentclass[floats,floatfix,showpacs,amsmath,amssymb,aps,twocolumn,superscriptaddress,nofootinbib,nolongbibliography,prd]{revtex4-1}

\usepackage{amssymb,amsmath,verbatim,mathtools,needspace,enumitem,etoolbox,graphicx,physics,microtype,afterpage,xspace,tabularx,lmodern,multirow}
\newcommand\numberthis{\addtocounter{equation}{1}\tag{\theequation}}
\usepackage{gensymb}
\usepackage{soul}
\usepackage{multirow}
\usepackage{subcaption}
\usepackage{ragged2e}
\usepackage{bbm}
\usepackage[dvipsnames, usenames]{xcolor}
\definecolor{linkcolor}{rgb}{0.0,0.3,0.5}
\usepackage[unicode, colorlinks=true, linkcolor=linkcolor, citecolor=linkcolor, filecolor=linkcolor, urlcolor=linkcolor, linktocpage, breaklinks]{hyperref}
\usepackage[all]{hypcap}
\usepackage{cancel}
\usepackage{empheq}
\usepackage[T1]{fontenc}
\usepackage[utf8]{inputenc}
\usepackage[usenames,dvipsnames]{xcolor}
\usepackage{csquotes}
\usepackage{mathrsfs}
\hypersetup{colorlinks=true,citecolor=Periwinkle,linkcolor=Periwinkle,urlcolor=Periwinkle}

\definecolor{romared}{RGB}{142,0,28}

\newcommand{\be}{\begin{equation}}
\newcommand{\ee}{\end{equation}}

\def\be{\begin{equation}}
\def\ee{\end{equation}}
\newcommand{\beq}{\begin{eqnarray}}
\newcommand{\eeq}{\end{eqnarray}}
\usepackage{makecell}
\usepackage{soul}

\newcolumntype{Y}{>{\centering\arraybackslash}X}

\newcommand{\vlambda}{\vec{\lambda}}
\newcommand{\vtheta}{\vec{\theta}}
\newcommand{\vpsi}{\vec{\psi}}
\newcommand{\vvarphi}{\vec{\varphi}}
\newcommand{\mathH}{\mathcal{H}}
\newcommand{\htheta}{\hat{\theta}}
\newcommand{\hpsi}{\hat{\psi}}

\newcommand{\vnorm}[1]{\left\lVert#1\right\rVert}

\makeatletter
\newcommand*{\rom}[1]{\expandafter\@slowromancap\romannumeral #1@}
\makeatother

\makeatletter
\let\jnl@style=\rm
\def\ref@jnl#1{{\jnl@style#1}}

\def\aj{\ref@jnl{AJ}}                   
\def\actaa{\ref@jnl{Acta Astron.}}      
\def\araa{\ref@jnl{ARA\&A}}             
\def\apj{\ref@jnl{ApJ}}                 
\def\apjl{\ref@jnl{ApJ}}                
\def\apjs{\ref@jnl{ApJS}}               
\def\ao{\ref@jnl{Appl.~Opt.}}           
\def\apss{\ref@jnl{Ap\&SS}}             
\def\aap{\ref@jnl{A\&A}}                
\def\aapr{\ref@jnl{A\&A~Rev.}}          
\def\aaps{\ref@jnl{A\&AS}}              
\def\azh{\ref@jnl{AZh}}                 
\def\baas{\ref@jnl{BAAS}}               
\def\bac{\ref@jnl{Bull. astr. Inst. Czechosl.}}
\def\caa{\ref@jnl{Chinese Astron. Astrophys.}}
\def\cjaa{\ref@jnl{Chinese J. Astron. Astrophys.}}
\def\icarus{\ref@jnl{Icarus}}           
\def\jcap{\ref@jnl{J. Cosmology Astropart. Phys.}}
\def\jrasc{\ref@jnl{JRASC}}             
\def\memras{\ref@jnl{MmRAS}}            
\def\mnras{\ref@jnl{MNRAS}}             
\def\na{\ref@jnl{New A}}                
\def\nar{\ref@jnl{New A Rev.}}          
\def\pra{\ref@jnl{Phys.~Rev.~A}}        
\def\prb{\ref@jnl{Phys.~Rev.~B}}        
\def\prc{\ref@jnl{Phys.~Rev.~C}}        
\def\prd{\ref@jnl{Phys.~Rev.~D}}        
\def\pre{\ref@jnl{Phys.~Rev.~E}}        
\def\prl{\ref@jnl{Phys.~Rev.~Lett.}}    
\def\pasa{\ref@jnl{PASA}}               
\def\pasp{\ref@jnl{PASP}}               
\def\pasj{\ref@jnl{PASJ}}               
\def\rmxaa{\ref@jnl{Rev. Mexicana Astron. Astrofis.}}%
\def\qjras{\ref@jnl{QJRAS}}             
\def\skytel{\ref@jnl{S\&T}}             
\def\solphys{\ref@jnl{Sol.~Phys.}}      
\def\sovast{\ref@jnl{Soviet~Ast.}}      
\def\ssr{\ref@jnl{Space~Sci.~Rev.}}     
\def\zap{\ref@jnl{ZAp}}                 
\def\nat{\ref@jnl{Nature}}              
\def\iaucirc{\ref@jnl{IAU~Circ.}}       
\def\aplett{\ref@jnl{Astrophys.~Lett.}} 
\def\apspr{\ref@jnl{Astrophys.~Space~Phys.~Res.}}
\def\bain{\ref@jnl{Bull.~Astron.~Inst.~Netherlands}} 
\def\fcp{\ref@jnl{Fund.~Cosmic~Phys.}}  
\def\gca{\ref@jnl{Geochim.~Cosmochim.~Acta}}   
\def\grl{\ref@jnl{Geophys.~Res.~Lett.}} 
\def\jcp{\ref@jnl{J.~Chem.~Phys.}}      
\def\jgr{\ref@jnl{J.~Geophys.~Res.}}    
\def\jqsrt{\ref@jnl{J.~Quant.~Spec.~Radiat.~Transf.}}
\def\memsai{\ref@jnl{Mem.~Soc.~Astron.~Italiana}}
\def\nphysa{\ref@jnl{Nucl.~Phys.~A}}   
\def\physrep{\ref@jnl{Phys.~Rep.}}   
\def\physscr{\ref@jnl{Phys.~Scr}}   
\def\planss{\ref@jnl{Planet.~Space~Sci.}}   
\def\procspie{\ref@jnl{Proc.~SPIE}}   

\makeatother

\usepackage{etoolbox} 
\makeatletter
\patchcmd{\@outputpage@head}{\@ifnum{\@mpcol+\@ne}{\@disablepaircolumn}{}}{}{}{}
\makeatother

\begin{document}

\title{Hierarchical modeling of gravitational-wave populations for disentangling environmental and modified-gravity effects}

\author{Shubham Kejriwal} 
\email[]{shubhamkejriwal@u.nus.edu}
\affiliation{Department of Physics, National University of Singapore, Singapore 117551}
\author{Enrico Barausse} 
\email[]{barausse@sissa.it}
\affiliation{SISSA, Via Bonomea 265, 34136 Trieste, Italy and INFN Sezione di Trieste}
\affiliation{IFPU - Institute for Fundamental Physics of the Universe, Via Beirut 2, 34014 Trieste, Italy}
\author{Alvin J. K. Chua}
\email[]{alvincjk@nus.edu.sg}
\affiliation{Department of Physics, National University of Singapore, Singapore 117551}
\affiliation{Department of Mathematics, National University of Singapore, Singapore 119076}

\begin{abstract}

   The upcoming Laser Interferometer Space Antenna (LISA)
    will detect up to thousands of extreme-mass-ratio inspirals (EMRIs). These sources will spend $\sim 10^5$ cycles in band, and are therefore sensitive to tiny changes in the general-relativistic dynamics, potentially induced by astrophysical environments or modifications of general relativity (GR). Previous studies have shown that these effects can be highly degenerate for a single source. However, it may be possible to distinguish between them at the population level, because environmental effects should impact only a fraction of the sources, while modifications of GR would affect all. We therefore introduce a population-based hierarchical framework to disentangle the two hypotheses. Using simulated EMRI populations, we perform  tests of the null vacuum-GR hypothesis and two alternative beyond-vacuum-GR hypotheses, namely migration torques (environmental effects) and time-varying $G$ (modified gravity). We find that with as few as $\approx 20$ detected sources, our framework can statistically distinguish between these three hypotheses, and even indicate if both environmental and modified gravity effects are simultaneously present in the population.  Our framework can be applied to other models of beyond-vacuum-GR effects available in the literature.
\end{abstract}

\maketitle 

\section{Introduction}

Since the first detection of gravitational waves (GWs) in 2015~\cite{LIGOScientific:2016aoc}, the LIGO-Virgo-KAGRA (LVK) GW Transient Catalog (GWTC) has grown to include more than 200 transient sources~\cite{LIGOScientific:2018mvr,LIGOScientific:2020ibl,LIGOScientific:2021usb,KAGRA:2021vkt,LIGOScientific:2025slb} in the $\sim 100$ Hz band. The GWTC has been used to set tight constraints on the properties of merging compact binary populations in the universe~\cite{LIGOScientific:2025pvj} and to test general relativity (GR)~\cite{LIGOScientific:2019fpa,LIGOScientific:2020tif,LIGOScientific:2021sio}. The upcoming network of third-generation ground-based detectors, including the Einstein Telescope~\cite{Hild:2010id} and Cosmic Explorer~\cite{LIGOScientific:2016wof,Reitze:2019iox}, will further improve these constraints and extend population studies to heavier binaries in the $\sim 1-10$ Hz band~\cite{Singh:2021zah,Ng:2020qpk,Gair:2010dx}. At even lower frequencies $\sim 1$ mHz, the upcoming Laser Interferometer Space Antenna (LISA) will probe the properties of supermassive black hole (MBH) populations~\cite{LISA:2024hlh, Gair:2010bx}. 

A particularly interesting class of sources in the LISA band is provided by extreme-mass-ratio inspirals (EMRIs), where a compact object (CO) of mass $\mu \sim 10^{1-2} M_\odot$ completes $\sim 10^5$ highly relativistic orbits around an MBH of mass $M \sim 10^{5-7} M_\odot$. EMRIs are expected to constrain MBH properties such as mass and spin to sub-percent precision~\cite{Babak:2017tow}, enabling tight constraints on the MBH mass function~\cite{Gair:2010yu}.
Furthermore, the large number of strong field cycles completed by the CO  makes EMRIs unique probes of astrophysical environments and potential deviations from GR~\cite{Barausse:2006vt, Barausse:2007dy,Kocsis:2011dr,Barausse:2014tra,Barausse:2016eii,LISA:2022kgy,Cole:2022yzw,Duque:2024mfw,Speri:2022upm,Speri:2024qak,Copparoni:2025jhq}. However, the sensitivity of EMRIs to  such ``beyond-vacuum-GR'' effects may hinder their precision astronomy prospects. These effects are secular and perturbative in nature, often manifesting as simple power-law-like corrections to the EMRI's vacuum-GR dynamics and introducing  correlations across the parameter space. If these effects are present in the signal but are not included in the analysis, they can  bias the source's inferred parameters and potentially lead to, e.g., false claims of deviations from GR~\cite{Speri:2022upm,Kejriwal:2023djc,Cardenas-Avendano:2024mqp}. 

Previous studies have employed hierarchical models to search for evidence of modified gravity effects in GW populations (see, e.g.,~\cite{Li:2011cg,Isi:2019asy,LIGOScientific:2019fpa, LIGOScientific:2020tif, LIGOScientific:2021sio, Liu:2023onj}), but do not consider multiple contending effects. Recently, Yuan et al.~\cite{Yuan:2024duo,Yuan:2025pbu} proposed a statistic $F$ to distinguish modified gravity effects (with a constant universal amplitude) from environmental effects when they are present in the signal \textit{one at a time}, but do not address the distinguishability of multiple concurrent effects in the system.

In this paper, we propose for the first time a framework to detect and distinguish between the two main classes of beyond-vacuum-GR effects using Bayesian hierarchical modeling, even when both classes of effects are present in the system simultaneously. The first class includes effects such as astrophysical environments, which introduce, e.g.,  corrections to the CO's orbit through tidal resonances~\cite{Bonga:2019ycj} or migration torques~\cite{Barausse:2007dy,Kocsis:2011dr,Barausse:2014tra,Speri:2022upm,Copparoni:2025jhq}. These effects can be independently realized across all sources, and may only be present in a fraction of the total population. We collectively refer to such effects as \textit{local} in this paper. The second class of effects includes fundamental physics modifications that are universal in nature, i.e., present in all sources simultaneously. In many cases, these corrections are controlled by an intrinsic ``strength'' constant across the population, such that GR is recovered when the strength parameter assumes a null value. An example is the time-varying Newton's constant $G$ (see e.g.~\cite{Yunes:2009bv, Chamberlain:2017fjl}), which yields GR when the rate of change of $G$ vanishes, i.e. $\dot{G} = 0$. In some other cases, e.g, in scalar-tensor theories~\cite{Brans:1961sx}, deviations from the GR predictions may also depend on the nature of the sources and/or its mass and spin~\cite{Sotiriou:2014pfa,Dima:2020yac,Maselli:2020zgv}, but are still expected to affect the whole EMRI population. We refer to this class of effects as \textit{global} in this paper.

This classification allows us to model local and global effects at different levels in a hierarchical framework. We can then perform hypothesis testing to compare the null vacuum-GR hypothesis with the local and global effect hypotheses, to test the preference of the data for one over another. Explicitly, the hypotheses that we will consider are: 
\begin{itemize}
    \item \textbf{Vacuum-GR hypothesis} \textbf{($\boldsymbol{\mathH_v}$)}: all sources evolve according to vacuum-GR dynamics;
    \item \textbf{Local effect hypothesis} \textbf{($\boldsymbol{\mathH_\ell}$)}: a nonzero fraction of the sources are affected by a local (e.g. astrophysical environment) effect, while the rest follow vacuum-GR;
    \item \textbf{Global effect hypothesis} \textbf{($\boldsymbol{\mathH_g}$)}: all sources experience a global (e.g. modified gravity) correction to the vacuum-GR dynamics.
\end{itemize}

\subsection{Executive summary of results}

In this paper, we apply the above framework to specific models of local and global effects. For the local effect, we choose migration torques, which are induced on the CO due to the presence of an accretion disk around the MBH. For the global effect, we select the time-varying $G$ model, with $\dot{G}~[\rm yr^{-1}]$ the constant intrinsic strength of the effect. We construct four different EMRI populations, $P_v$, $P_\ell$, $P_g$, and $P_{\rm mix}$. The first three, $P_v$, $P_\ell$, and $P_g$, are generated assuming $\mathH_v$, $\mathH_\ell$, and $\mathH_g$ as the truth, respectively. The fourth ``mixed'' population $P_{\rm mix}$ is constructed such that all sources experience a time-varying $G$, while a fraction of them are simultaneously affected by migration torques. We also consider three different population sizes, with 100, 500, and 1000 sources, of which $\approx 17\%$ are observable. We find that even for the population with 100 sources (of which $17$ are observable), the hierarchical framework can reliably recover the correct hypothesis under which the population was constructed, and even indicate the concurrent presence of local and global effects in the population.

We also gauge the robustness of our method for different choices of local and global effect parameters in the true population. We find that if the fraction $f$ of the population with a local effect is $\geq 0.9$, the recovered intrinsic strength of time-varying $G$, i.e. $\dot{G}~[\rm yr^{-1}]$, may be significantly biased. Furthermore, we highlight a special case where both local and global effects are present in the population simultaneously, but they effectively ``cancel'' each other out. In such unique scenarios, the log-Bayes factor $\log\mathcal{B}^g_\ell$, which quantifies the data's preference for the global effect hypothesis over the local one, is $\approx 0.0$, and the hierarchical framework struggles to distinguish between the two effects. The results are presented in more detail in Sec.~\ref{sec:results}.

\subsection{Structure of the paper}

The remainder of this paper is organized as follows. In Sec.~\ref{sec:background}, we begin with an overview of Bayesian hierarchical modeling and hypothesis testing for GW data analysis; In Sec.~\ref{sec:setupforgw}, we describe our source and population models under each of the three hypotheses, $\mathH_v$, $\mathH_\ell$, and $\mathH_g$; In Sec.~\ref{sec:dataanalysis}, we specify our data analysis setup, including a description of the approximate methods used for feasible hypothesis testing in the hierarchical framework; In Sec.~\ref{sec:results}, we present our results, applying the hierarchical framework to various examples of EMRI populations; Finally, in Sec.~\ref{sec:discussion}, we summarize our results, discuss their impact, and lay out potential future directions. 

Throughout this paper, we assume a flat universe with Hubble's constant $H_0 = 70$ km/s/Mpc, the matter density $\Omega_{m,0} = 0.3$, and the dark energy density $\Omega_{\Lambda,0} = 0.7$.

\section{Background}\label{sec:background}
\subsection{Bayesian hierarchical modeling}

Consider a population of $N_{\rm pop}$ GW sources modeled by a set of hyperparameters $\vlambda$. Each source is modeled by a set of parameters $\vtheta$, distributed according to some prior probability density function (pdf), $p(\vtheta|\vlambda, \mathH)$. Here, $\mathH$ is the assumed hypothesis. Of these $N_{\rm pop}$ sources, only a subset of size $N_{\rm obs} < N_{\rm pop}$ will satisfy a chosen detectability criterion, such that they are considered to be ``observed''. The measured data $\vec{d}_i$ of the $i^{\rm th}$ observed sources is represented by its maximum-likelihood estimate (MLE), $\htheta_i$,  where $i = 1, \ldots, N_{\rm obs}$~\cite{Mandel_2019}.\footnote{This setup, based on~\cite{Mandel_2019}, assumes that each $i^{\rm th}$ source has its own datastream $\vec{d}_i$, which does not directly apply to the LISA context, where all sources will overlap in the datastream instead. We can still apply this framework if we assume $\vec{d}_i$ to be the \textit{residual} of the $i^{\rm th}$ source when all other sources are subtracted from the full datastream, i.e., $\vec{d}_i = \vec{d}-\sum_{j\neq i} \vec{h}(\htheta_j)$. Here, $\vec{d}$ is the full LISA datastream and $\vec{h}(\htheta_j)$ is the single-source GW template evaluated at $\htheta_j$.} At the MLE point, the individual source likelihood $p(\vec{d}_i|\vtheta_i,\mathH)$ maximizes.

Given the set of measured data for the detected sources $\{\vec{d}_i\}$, the (hyper)posterior pdf $p(\vlambda|\{\vec{d}_i\},\mathH)$ of the hyperparameter set $\vlambda$ can be written according to Bayes theorem as~\cite{gelman2013bayesian}
\begin{align}
    p(\vlambda|\{\vec{d}_i\},\mathH) = \frac{p(\{\vec{d}_i\}|\vlambda,\mathH)\pi(\vlambda|\mathH)}{\mathcal{Z}(\{\vec{d}_i\}|\mathH)}.\label{eq:bayestheorem}
\end{align}
Here $p(\{\vec{d}_i\}|\vlambda,\mathH)$ is the hyperlikelihood of $\vlambda$, $\pi(\vlambda|\mathH)$ is its hyperprior, and $\mathcal{Z}(\{\vec{d}_i\}|\mathH)$ is the evidence for the data under the hypothesis, written explicitly as 
\begin{align}
    \mathcal{Z}(\{\vec{d}_i\}|\mathH) = \int {\rm d}\vlambda~ p(\{\vec{d}_i\}|\vlambda,\mathH)\pi(\vlambda|\mathH).\label{eq:evidence}
\end{align}
Assuming that individual sources are realized independently in the population, and using the correspondence between $\vec{d}_i$ and $\htheta_i$, we can write the hyperlikelihood as an effective product of single-source hyperlikelihoods~\cite{Mandel_2019}
\begin{align}
    p(\{\vec{d}_i\}|\vlambda,\mathH) = \prod_{i=1}^{N_{\rm obs}}\frac{p(\htheta_i|\vlambda,\mathH)}{\alpha(\vlambda|\mathH)},\label{eq:hyperlikelihood}
\end{align}
where $\alpha(\vlambda|\mathH)$ encodes the selection biases in the observation of these sources. With this, the hyperposterior of $\vlambda$ can be rewritten as
\begin{align}
    p(\vlambda|\{\vec{d}_i\},\mathH) = \frac{\pi(\vlambda|\mathH)}{\mathcal{Z}(\{\vec{d}_i\}|\mathH)}\times\prod_{i=1}^{N_{\rm obs}}\frac{p(\htheta_i|\vlambda,\mathH)}{\alpha(\vlambda|\mathH)}.~\label{eq:hyperposterior}
\end{align}

\subsection{Hypothesis testing}

We are interested in testing the vacuum-GR hypothesis $\mathH_v$ against the alternate beyond-vacuum-GR hypothesis $\mathH_a$ where $a = \ell, g$. These hypotheses can be modeled in a nested framework, with $\mathH_v \subseteq \mathH_a$. The equality holds when a general subset of hyperparameters $\vlambda' \subseteq \vlambda$ under $\mathH_a$ assumes a fiducial null value $\vlambda'_0$. Then, we can write the Bayes factor for the data's preference of $\mathH_v$ over $\mathH_a$ as the Savage-Dickey ratio~\cite{gelman2013bayesian,10.1214/aoms/1177693507},
\begin{align}
    \mathcal{B}_a^v = \frac{p(\vlambda'=\vlambda'_0|\{\vec{d}_i\},\mathH_a)}{\pi(\vlambda'=\vlambda'_0|\mathH_a)}; ~a = \ell,g.~\label{eq:SDgeneral}
\end{align}
Here, the numerator is the marginalized hyperposterior of $\vlambda'$ at $\vlambda'_0$, and the denominator is the corresponding marginalized hyperprior. Substituting the hyperposterior~\eqref{eq:hyperposterior} and the evidence~\eqref{eq:evidence}, we can rewrite $\mathcal{B}^v_a$ as
\begin{align}
    \mathcal{B}^v_a = \frac{\prod_i p(\htheta_i|\vlambda'=\vlambda'_0,\mathcal{H}_a)}{\int{\rm d}\vlambda'~\prod_ir_\alpha ~p(\htheta_i|\vlambda',\mathcal{H}_a)\pi(\vlambda'|\mathH_a)}~\label{eq:SDsemifinal}
\end{align}
where we define $r_{\alpha} := \alpha(\vlambda'=\vlambda'_0|\mathH_a)/\alpha(\vlambda'|\mathH_a)$. 

In the following, we set $r_\alpha = 1$, effectively treating $\alpha(\vlambda'|\mathH_a)$ as a constant. As argued more formally in Appendix~\ref{app:constmargselfun}, this assumption holds when $\vlambda'$ is a set of beyond-vacuum-GR effect hyperparameters, e.g., the intrinsic strength of time-varying $G$. This is because such effects only perturbatively modify the detectability of a source in the population. Setting $r_\alpha = 1$ mitigates some computational expense in the calculation of the Bayes factor, as the selection function involves an integral over source parameters over the full space. The integrand includes the detection probability $p_{\rm det}(\vtheta|\mathH)$, which is typically a numerical function in the GW context and hence expensive to compute. If $p_{\rm det}(\vtheta|\mathH)$ can be cheaply calculated, this factor may be restored feasibly in future studies (see, e.g.,~\cite{Chapman-Bird:2022tvu,Singh:2025pzh}). 

In Eq.~\eqref{eq:SDsemifinal}, the single source hyperlikelihood $p(\htheta_i|\vlambda',\mathH_a)$ can be expressed as a marginal over $\vtheta$,
\begin{align}
    p(\htheta_i|\vlambda',\mathH_a) = \int {\rm d}\vtheta~p(\htheta_i|\vtheta,\mathH_a)\pi(\vtheta|\vlambda',\mathH_a).~\label{eq:sourceintegral}
\end{align}
Here, under the hypothesis $\mathH_a$, $p(\htheta_i|\vtheta,\mathH_a)$ is the likelihood of $\vtheta$ under the data $\htheta_i$, $\pi(\vtheta|\vlambda',\mathH_a)$ is the prior pdf of $\vtheta$ given $\vlambda'$ used in the analysis, and the integral is over the full space of $\vtheta$. While this explicit integral evaluation can make hypothesis testing expensive, we can analytically approximate it for the models of our interest, as we shall describe in Sec.~\ref{sec:approximatehyperlike}. 

For simplicity, we additionally assume flat priors on the model hyperparameters $\vlambda$, such that $\pi(\vlambda'|\mathH_a) = \pi(\vlambda'=\vlambda'_0|\mathH_a)$. Hence, the final form of the Bayes factor used for hypothesis testing under our assumptions is
\begin{align}
    \mathcal{B}^v_a = \frac{1}{\pi(\vlambda'=\vlambda'_0|\mathH_a)}\frac{\prod_i p(\htheta_i|\vlambda'=\vlambda'_0,\mathcal{H}_a)}{\int{\rm d}\vlambda'\prod_ip(\htheta_i|\vlambda',\mathcal{H}_a)}.~\label{eq:savagedickeyratio}
\end{align}

\section{Setup}\label{sec:setupforgw}

\subsection{Source model}~\label{sec:constructingpopulation}

In this paper, we model individual sources as Kerr EMRIs evolving in circular and equatorial orbits. Our model allows for generic corrections to the vacuum-GR dynamics from migration torques and time-varying $G$ effects. We model these effects as simple power-law corrections to the leading-order GR flux of the axial component of the angular momentum~\cite{Speri:2022upm},
\begin{align}
    \dot{L} = \dot{L}_{\rm GR}\left(1 + A_{\ell}\left(\frac{p}{10M}\right)^{n_{\ell}}  + A_{g}\left(\frac{p}{M}\right)^{n_{g}} \right).\label{eq:angularmomentumbvgr}
\end{align}
Here $p$ is the instantaneous semi-latus rectum of the CO's orbit and $M$ is the MBH mass; $A_\ell$ is the effective strength of the migration torque effect, with typical values $\sim 10^{-5}$~\cite{Barausse:2014tra}; $n_\ell$ is the post-Newtonian (PN) order at which the local effect enters the evolution, and is $\approx 8$ for migration torques in $\alpha$-disks~\cite{Barausse:2014tra}; $A_g$ is the strength of the time-varying $G$ effect, with current best constraints $\lesssim 10^{-12}~[\rm yr^{-1}]$~\cite{Chamberlain:2017fjl}, and $n_g$ is its PN order in the evolution. For time-varying $G$, $n_g = 4$, denoting a $-4$ PN order correction~\cite{Chamberlain:2017fjl,Speri:2022upm}. Note that the flux balance equation reduces to vacuum-GR as $A_\ell, A_g \to 0$. We emphasize that such parametric constructions are available for other beyond-vacuum-GR effects in the literature~\cite{Kocsis:2011dr,Barausse:2014tra,Barausse:2016eii,LISA:2022kgy,Cole:2022yzw,Duque:2024mfw,Speri:2024qak,Kejriwal:2023djc}.

In addition to those described above, several other parameters define the EMRI model: $a$, the dimensionless spin of the black hole; $\mu$, the mass of the CO; $d_L$, the luminosity distance of the source, or equivalently the redshift $z \equiv z(d_L)$; $(\theta_S,\phi_S)$, the sky localization angles; $(\theta_K, \phi_K)$, the spin orientation angles; and $\Phi_0$, the initial azimuthal orbital phase. 

When inferring sources in the population, we only consider a subset of these parameters. We choose two vacuum-GR parameters, $\vec{v} := \{\ln M, z\}$, two local effect parameters, $\vec{\ell} := \{A_\ell, n_\ell\}$, and one global effect parameter $\vec{g} := \{A_g\}$. Hence we can define the source parameter set $\vtheta := \{\vec{v}, \vec{\ell}, \vec{g}\}$ for data analysis.

\subsection{Population model}

Given the source model, the EMRI population can be described by $N_{\rm pop}$ draws of the source parameters $\vtheta \sim \pi(\vtheta|\vlambda,\mathH)$. We decompose the hyperparameter set similarly to the source parameters, $\vlambda := \{\vlambda_v,\vlambda_\ell,\vlambda_g\}$, and assume independent priors on $\vec{v}$, $\vec{\ell}$, and $\vec{g}$, such that the joint prior $\pi(\vtheta|\vlambda,\mathH)$ can be written piecewise,
\begin{align}
\pi(\vtheta|\vlambda,\mathH) = \pi(\vec{v}|\vlambda_v)\times\begin{cases}  1 & \mathH = \mathH_v\\
\pi(\vec{\ell}|\vlambda_\ell) & \mathH = \mathH_\ell\\
\pi(\vec{g}|\vlambda_g) & \mathH=\mathH_g
\end{cases}.~\label{eq:priordecomposed}
\end{align}

For the vacuum-GR parameters $\vec{v} = \{\ln M, z\}$, we choose a prior pdf $p(\vec{v}|\vlambda_v)$ that scales with the MBH mass $M$ and the source redshift $z$, and also accounts for the size of the comoving volume. Explicitly, we choose (see e.g.~\cite{Gair:2010yu})
\begin{align}
    \pi(\vec{v}|\vlambda_v) &\propto d_c^2(z)\frac{{\rm d}n}{{\rm d}\ln{M}{\rm d}z},~\label{eq:priorvacuum}\\
    \frac{{\rm d}n}{{\rm d}\ln{M}{\rm d}z} &:= K\left(\frac{M}{M_*}\right)^{\alpha}(1+z)^\beta .
\end{align}
Here, the comoving distance $d_c(z)$ is related to the redshift $z$ given $H_0$, $\Omega_{m,0}$, and $\Omega_{\Lambda,0}$~\cite{Baumann:2022mni}. Additionally, when calculating the number density of sources, ${\rm d}n/{\rm d}\ln M {\rm d}z$, $K$ determines the size of the population, and $M_*$ sets the scale for the MBH masses. The corresponding hyperparameter set is $\vlambda_v := \{K,\alpha,\beta\}$, and we fix $M_* = 3\times 10^6 M_\odot$, following Babak et al.~\cite{Babak:2017tow}.

The local effect parameters $\vec{\ell} = \{A_\ell, n_\ell\}$ must be drawn from a distribution such that only a fraction $:= f$ of the sources in the population have a non-null value of $\vec{\ell}$, while the remaining $(1-f)$ sources follow a vacuum-GR evolution. To incorporate this dependence on $f$, we choose a bimodal distribution for the prior,
\begin{align}
    \pi(\vec{\ell}|\vlambda_\ell):=(1-f)\delta^2(\vec{\ell}) + f\mathcal{N}(\vec{\mu}_\ell|\vec{\ell},\Sigma_\ell).~\label{eq:priorlocal}
\end{align}
With this choice, a fraction $f$ of the sources is chosen from a two-dimensional normal distribution $\mathcal{N}(\vec{\mu}_\ell|\vec{\ell},\Sigma_\ell)$ with a mean $\vec{\mu}_\ell := (\mu_{A_\ell}, \mu_{n_\ell})$ and covariance $\Sigma_\ell := {\rm diag}(\sigma_{A_\ell}^2, \sigma_{n_\ell}^2)$. The remaining $(1-f)$ draws are identically zero, i.e., they assume the null value of $A_\ell = n_\ell = 0$, enforced by the Dirac-delta $\delta^2(\vec{\ell})$. Correspondingly, the local effect hyperparameter set is $\vlambda_\ell := \{f, (\mu_{A_\ell},\mu_{n_\ell}),(\sigma_{A_\ell},\sigma_{n_\ell})\}$.

Finally, for the global effect, we only model the intrinsic strength of time-varying $G$ with $\vec{g} = \{A_g\}$. Since this strength is modeled to be constant in all sources, we choose a Dirac-delta prior,
\begin{align}
    \pi(\vec{g}|\vlambda_g):= \delta(A_g-\dot{G})~\label{eq:priorglobal}
\end{align}
where $\dot{G}~[\rm yr^{-1}]$ is the value of time-varying $G$ at the population level. Hence, the global effect hyperparameter $\vlambda_g = \{\dot{G}\}$. With this definition, while the inferred source parameter $A_g$ and the hyperparameter $\dot{G}$ are equivalent in the signal model, we stress that they may assume different values in the analysis. For example, in the global effect hypothesis, the inferred MLE of $A_g$ will be systematically biased by the true value of the local effect (migration torque) parameters in a given source. Since the local effect's value changes across sources, the MLE $\hat{A}_g$ also becomes a source-dependent quantity. On the other hand, the inferred best-fit value of the hyperparameter $\dot{G}$, which is constant for all sources by definition, is inferred by ``averaging'' over these biased, source-dependent estimates of $A_g$. This averaging corresponds to the product over individual source likelihoods in the definition of the hyperposterior (Eq.~\eqref{eq:hyperposterior}).

\section{Data analysis techniques}\label{sec:dataanalysis}

Assuming a fiducial true hypothesis $\mathH^*$, and a corresponding set of true hyperparameters $\vlambda^*$, we can now construct a population of $N_{\rm pop}$ sources with inferred parameters $\{\vtheta^*\}$ drawn from the prior $\pi(\vtheta|\vlambda^*,\mathH^*)$. Here, we formulate an approximate analysis model to infer these parameters, which can then be used to perform hypothesis testing. For simplicity, we assume the zero-detector-noise realization in this paper, such that our results may be interpreted as averages over many noise realizations drawn from a zero-mean Gaussian process.

\subsection{Obtaining the MLE estimate of observed sources}

To claim an observation, we define a detectability criterion based on the optimal signal-to-noise ratio (SNR) of the signal. Explicitly, we set the criterion $\rho^2_{\rm opt} := \langle \vec{h}(\vtheta^*)|\vec{h}(\vtheta^*)\rangle \geq \rho^2_{\rm thresh}$. Here $\langle \cdot |\cdot \rangle$ is the detector-noise-weighted inner-product~\cite{Finn:1992wt, Cutler:1994ys}, and $\vec{h}$ is the GW template evaluated at $\vtheta^*$. $\rho_{\rm thresh}$ is the fiducial choice of the SNR threshold, which we will quantify below.

The value of the MLE $\htheta_i$ for the $i^{\rm th}$ observed source depends on the hypothesis assumed in the analysis. If the correct hypothesis $\mathH^*$ is assumed, the MLE is simply $\htheta_i = \vtheta_i^*$. Otherwise, the MLE will be systematically biased away from the truth~\cite{Cutler:2007mi,Cutler:1994ys,Flanagan:1997kp}. Assuming the linearized-signal approximation (LSA)~\cite{Vallisneri:2007ev}, we can approximate these biases by the linear-bias approximation for nested models~\cite{Kejriwal:2023djc}. We provide its explicit form in Eq.~\ref{eq:CVnested} of the Appendix~\ref{app:linearbiasapprox}. 

Since the LSA (and hence the linear-bias approximation) holds for ``large'' SNRs, we can impose that it is valid for the observed sources by setting a similarly large $\rho_{\rm thresh}$. We choose $\rho_{\rm thresh} = 20$, which is commonly chosen in the literature (see, e.g.,~\cite{Finn:1992wt,Cutler:1994ys,Vallisneri:2007ev}).

In realistic inference setups, the employed sampling scheme may be limited to a small region of the parameter space for efficiency. Thus, given a hypothesis, any source with an MLE (as estimated by the linear-bias approximation) that lies beyond this region will effectively remain undetected. To include this effect in our analysis, we set fiducial lower and upper bounds $[a_k, b_k]$ on all $k = 1, \ldots, D$ inferred model parameters, and omit any sources in the observed population for which the MLE is outside these bounds.\footnote{When the local effect amplitude $A_\ell$ is small, the linear-bias approximation may predict the slope $n_\ell$'s MLE estimate to be significantly beyond any realistic bounds, due to its near-zero information content. Thus, for small $A_\ell$, we will exempt $n_\ell$ from the inference bounds rule (in such cases, $n_\ell$ is uninformative anyway and hence does not contribute to the hyperlikelihood's value).} We describe the chosen bounds for our analysis in Table~\ref{tab:inferencebounds}. In practice, we found that reasonably changing these bounds does not significantly affect our results.

\begin{table}[]
    \centering
    \begin{tabular}{c c c}
        \hline
        \multirow{2}{*}{\textbf{parameter}} & \multirow{2}{*}{\textbf{lower bound}} & \multirow{2}{*}{\textbf{upper bound}} \\
         & & \\
        \hline
        \hline
        & & \\
        $\ln M$ & $\log(10^{5.5})$ & $\log(10^{6.5})$ \\
        $z$ & $0.01$ & $1.0$ \\
        $A_l$ & $0.0$ & $10^{-5}$\\
        $n_l$ & $-20.0$ & $20.0$\\
        $A_g$ & $-5\times 10^{-12}$ & $5\times 10^{-12}$\\
        & & \\
        \hline
    \end{tabular}
    \caption{\justifying Inference bounds on the source parameters. Given a hypothesis, any source in the true population with an MLE that lies beyond these bounds effectively remains undetected.}
    \label{tab:inferencebounds}
\end{table}

\subsection{Approximate hyperlikelihoods}~\label{sec:approximatehyperlike}

To evaluate the Bayes factor in Eq.~\eqref{eq:savagedickeyratio}, we must calculate the full-space integral over the source parameters $\vtheta$~\eqref{eq:sourceintegral} to obtain the single source hyperlikelihood~\eqref{eq:hyperlikelihood} in the alternative hypothesis. In general, evaluating this integral can be computationally expensive or practically infeasible. This is especially true in realistic inference setups in GW data analysis, which involve a high-dimensional parameter space. 

With the framework constructed above, we show in Appendices~\ref{app:vacuumhyperlikelihoodapp},~\ref{app:localhyperlikelihoodapp}, and~\ref{app:globalhyperlikelihoodapp} that Eq.~\eqref{eq:sourceintegral} can be analytically approximated under each of the three hypotheses $\mathH_v$, $\mathH_\ell$, and $\mathH_g$, respectively. Our results make use of the LSA to approximate the individual source likelihood in a given hypothesis $\mathH$ by a normal distribution,
\begin{align}
    p(\htheta_i|\vtheta,\mathH) \approx \mathcal{N}(\vtheta|\htheta_i,\Gamma_i^{-1}).~\label{eq:normallikelihood}
\end{align}
Here, $\Gamma_i$ is the Fisher information matrix (FIM), with elements
\begin{align}
    (\Gamma_i)_{km} := \langle\partial_kh\big|\partial_mh\rangle\big|_{\htheta_i} \approx \langle \partial_k h|\partial_m h\rangle \big|_{\vtheta_i^*}
\end{align} 
and $\partial_kh := \partial h/\partial\vtheta^k$.

We also validate the constructed approximate hyperlikelihoods against the direct Monte Carlo integration of Eq.~\eqref{eq:sourceintegral}. We describe the validation scheme in detail in Appendix~\ref{app:validation}. As shown in Fig.~\ref{fig:integralvalidation} of the Appendix, the results are consistent up to stochastic errors in the Monte Carlo integration.

\section{Results}~\label{sec:results}

With the hierarchical framework developed in Sec.~\ref{sec:background}, the description of the EMRI model in Sec.~\ref{sec:setupforgw}, and the data analysis setup in Sec.~\ref{sec:dataanalysis}, we now perform hypothesis tests of beyond-vacuum-GR effects in various example cases. In the first subsection, we study how well our framework can recover the true hypothesis under which a given EMRI population was generated, considering three different population sizes. Then, in the second subsection, we show the dependence of the Bayes factors on the true local and global effect hyperparameters $f^*$ and $\dot{G}^*$ in the population. This will allow us to comment on the robustness of our framework, particularly in disentangling these effects. Finally, we briefly comment on the distinguishability of other beyond-vacuum-GR effects within the hierarchical framework. All results presented below are produced using the \texttt{Hierarchical} package, publicly available on Github (\url{https://github.com/perturber/Hierarchical/}).

\subsection{Fixed population studies}~\label{sec:populationanalysis}

\subsubsection{Setup}

\noindent{\textit{Population and source parameters ---}} We construct three EMRI populations $P_v$, $P_\ell$, and $P_g$, generated assuming $\mathH_v$, $\mathH_\ell$, and $\mathH_g$, as the true hypothesis, respectively. In addition, we consider a fourth population $P_{\rm mix}$, in which the (global) time-varying $G$ effect is present in all sources and a fraction $f$ of the population is simultaneously affected by (local) migration torques. The true hyperparameters in all four populations are given in Table~\ref{tab:truehyper} in the Appendix. 

In $P_\ell$ and $P_{\rm mix}$, we choose the mean of the local effect amplitude and slope as $\mu_{A_\ell}, \mu_{n_\ell} = (10^{-6}, 8)$. $\mu_{n_\ell}$ is chosen as the value of the migration torque slope for an $\alpha$-disk model and $\mu_{A_\ell}$ is chosen to be an order-of-magnitude smaller than typical values of the effective migration torque amplitudes $\sim 10^{-5}$~\cite{Barausse:2014tra}. We choose a smaller value to ensure that the linear-bias approximation, valid for small corrections, conservatively holds for all generated sources. The fractional size of the population with the local effect is fixed to $f=0.5$. In $P_g$ and $P_{\rm mix}$, the global effect hyperparameter $\dot{G}^*$ is chosen to be $10^{-12} [{\rm yr}^{-1}]$. This value is consistent with current bounds on $\dot{G} \lesssim 10^{-12} [\rm yr^{-1}]$~\cite{Chamberlain:2017fjl}.

Given the true hyperparameters, we consider EMRI populations of three sizes, with $N_{\rm pop} = $ 100, 500, and 1000 sources. The inferred source parameters $\vec{v} = \{\ln M, z\}$, $\vec{\ell} = \{A_\ell, n_\ell\}$, and $\vec{g} = \{A_g\}$ are drawn from their respective prior pdfs, i.e. Eqs.~\eqref{eq:priorvacuum},~\eqref{eq:priorlocal}, and~\eqref{eq:priorglobal}. Since the priors on $\ln M, z$ are improper, we additionally set minimum and maximum cutoffs on their ranges. We choose $M \in [10^{5.5}, 10^{6.5}]$, and $z \in [0.01, 1.0]$. The luminosity distance, $d_L(z)$, of the source is computed from the drawn redshifts $z$ for the assumed values of $H_0$, $\Omega_{m,0}$, and $\Omega_{\Lambda,0}$. 

The remaining (uninferred) source parameters, $\mu, a, p_0, \theta_S, \phi_S, \theta_K, \phi_K, \Phi_0, n_g$, are drawn from some fiducial priors, as summarized in Table~\ref{tab:priorsall} in the Appendix. All except $p_0$ and $n_g$, i.e., the initial semi-latus rectum of the CO's orbit and the global effect slope, respectively, are drawn uniformly. $p_0$ is set to $p_{0}(T_{\rm plunge})$ where the CO with an initial semi-latus rectum $p_{0}(T_{\rm plunge})$ plunges into the MBH after a randomly chosen time $T_{\rm plunge} \sim \mathcal{U}[0.5, 2]$ years; $n_g = 4$ is fixed in all sources because the chosen global effect, i.e., time-varying $G$ is a -4 PN order effect~\cite{Chamberlain:2017fjl,Speri:2022upm}; We thus obtain the parameter set $\{\vtheta^*\}$ of all $N_{\rm pop}$ sources in the population.

\noindent{\textit{Detector response and GW observables ---}} For a given source with true parameters $\vtheta^*$, we generate a time-domain GW waveform $\vec{h}(\vtheta^*) := \vec{h}_+(\vtheta^*) - i\vec{h}_\times(\vtheta^*)$ in the solar system barycenter (SSB) frame. We use the adiabatic Kerr equatorial inspiral model \texttt{FastKerrEccentricEquatorial} of the \textsc{FastEMRIWaveforms (FEW)} package~\cite{Chua:2020stf,Katz:2021yft,Speri:2023jte,Chapman-Bird:2025xtd}. Additionally, we fix the detector's observation window to 1 year, such that sources that plunge earlier are observed until merger. Otherwise, their waveforms are cut off at the 1-year mark. 

This SSB frame waveform is then modulated by the LISA response function. Working in the 1st-generation time-delay interferometry (TDI) configuration with equal armlengths, we can express the GW signal as three TDI observables, $A$, $E$, and $T$~\cite{Vallisneri:2004bn,Dhurandhar:2001tct}. However, the $T$ channel can be dropped because its GW strain content is negligible compared to the $A$ and $E$ channels in this configuration.\footnote{Remember that we assumed a zero detector-noise realization in Sec.~\ref{sec:constructingpopulation} such that the noise information in the $T$ channel is also zero.} We use the \texttt{FastLISAResponse} module of the \textsc{lisa-on-gpu} package~\cite{Katz:2022yqe} to obtain the $A$, $E$ observables $:= \vec{h}_{A,E}(\vtheta^*)$.

\noindent\textit{Detected sources and MLE estimates---} Of the $N_{\rm pop}$ GW signals generated above, only $N_{\rm obs} < N_{\rm pop}$ will be observed. They satisfy the detectability criterion (Sec.~\ref{sec:dataanalysis})
\begin{align}
\rho^2_{\rm opt} = \sum_{C=A,E}\langle\vec{h}_{C}(\vtheta^*)|\vec{h}_{C}(\vtheta^*)\rangle \geq \rho^2_{\rm thresh}. 
\end{align}
This constitutes the set of $i = 1, \ldots, N_{\rm obs}$ observed sources in the population. For the $i^{\rm th}$ source, we calculate the FIM in the full parameter space, $\Gamma_i$, using the \textsc{StableEMRIFisher (SEF)} package~\cite{kejriwal_2024_sef}. The corresponding MLE estimate, $\htheta_i$, is obtained assuming the linear-bias approximation, Eq.~\eqref{eq:CVnested}.

With the set of MLEs $\{\htheta_i\}$ in hand, we can now calculate the Bayes factor from Eq.~\eqref{eq:savagedickeyratio}. Explicitly, the Bayes factors comparing the null vacuum-GR hypothesis against the alternative local effect and global effect hypotheses are, respectively,
\begin{align}
    \mathcal{B}^v_\ell &= \frac{1}{\pi(f=0|\mathH_\ell)}\frac{\prod_i p(\htheta_i|f=0,\mathH_\ell)}{\int {\rm d}f\prod_ip(\htheta_i|f,\mathH_\ell)},\label{eq:MCbayesfactorlocal}\\
    \mathcal{B}^v_g &= \frac{1}{\pi(\dot{G}=0|\mathH_g)}\frac{\prod_i p(\htheta_i|\dot{G}=0,\mathH_g)}{\int{\rm d}\dot{G}\prod_ip(\htheta_i|\dot{G},\mathH_g)}\label{eq:MCbayesfactorglobal}.
\end{align}
In practice, the integral in the denominator of these expressions can be approximated by a Monte Carlo integral with $N$ random draws of $f, \dot{G}$ from their respective hyperpriors $\pi(\vlambda|\mathH)$. Our choice of hyperpriors for all modeled hyperparameters is summarized in Table~\ref{tab:hyperpriorsall}. The Bayes factor for the global effect hypothesis over the local one is given simply by the ratio $\mathcal{B}^g_\ell = \mathcal{B}^v_\ell/\mathcal{B}^v_g$.

\subsubsection{Results}

\begin{figure}
    \centering
    \includegraphics[width=0.9\linewidth]{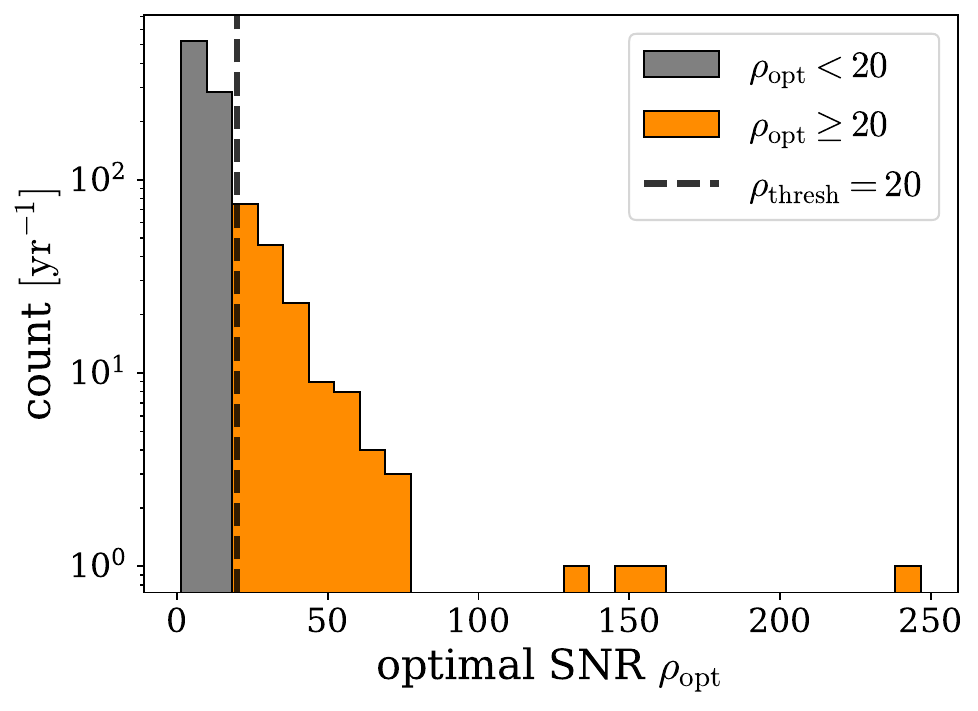}
    \caption{\justifying The distribution of optimal SNR $\rho_{\rm opt}$ in the vacuum-GR population $P_v$ for the $N_{\rm pop} = 1000$ case. SNRs $\rho_{\rm opt} \geq 20$ are highlighted in orange, representing $\approx 17\%$ of all sources.}
    \label{fig:SNRdist}
\end{figure}

\begin{figure*}
    \centering
    \begin{subfigure}[t]{0.9\textwidth}
        \centering
        \includegraphics[width=\textwidth]{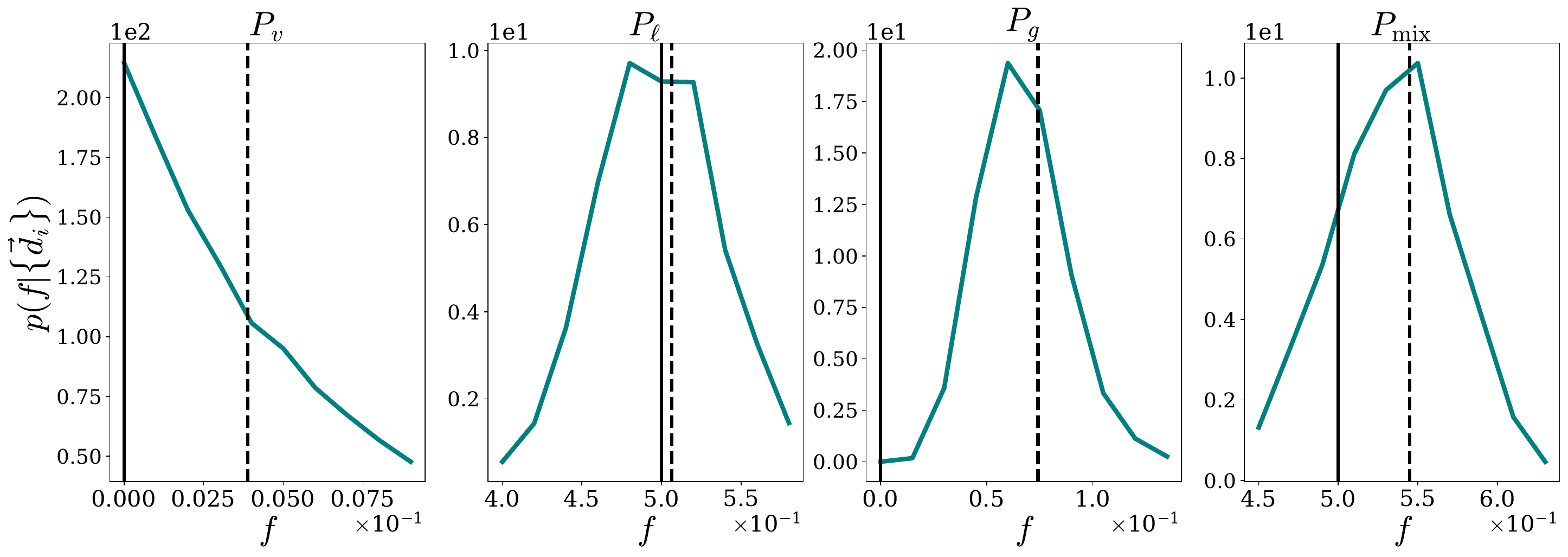} 
    \end{subfigure}
    \begin{subfigure}[t]{0.9\textwidth}
        \centering
        \includegraphics[width=\textwidth]{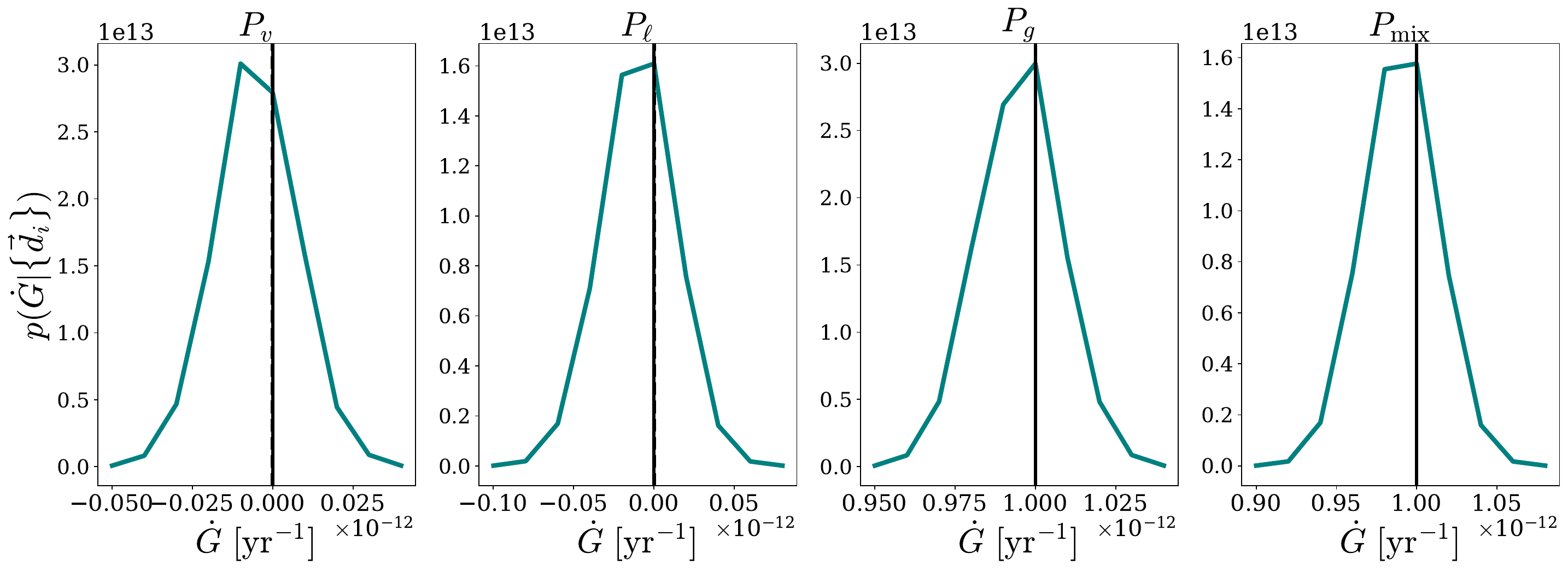}
    
    \end{subfigure}
    \caption{\justifying Marginalized hyperposteriors on $f$ (top panels) and $\dot{G}$ (bottom panels) constructed from $N = 5000$ samples from the analysis priors weighted by the hyperposterior densities. We plot four populations, namely, $P_v$, $P_\ell$, $P_g$, and $P_{\rm mix}$ (left to right) with population size $N_{\rm pop} = 1000$. The solid vertical line in each plot represents the true value of $f$ or $\dot{G}$ while the dashed vertical line represents the sample expectation. $f$ is found to be systematically biased by $\sim 3.5\sigma$ in $P_g$ and $\sim 1.5\sigma$ in $P_{\rm mix}$, while $\dot{G}$ is recovered consistently with the truth at $\lesssim 0.01\sigma$ in all cases.}
    \label{fig:marginalhyperlike}
\end{figure*}

\begin{figure*}
    \centering
    \includegraphics[width=0.9\linewidth]{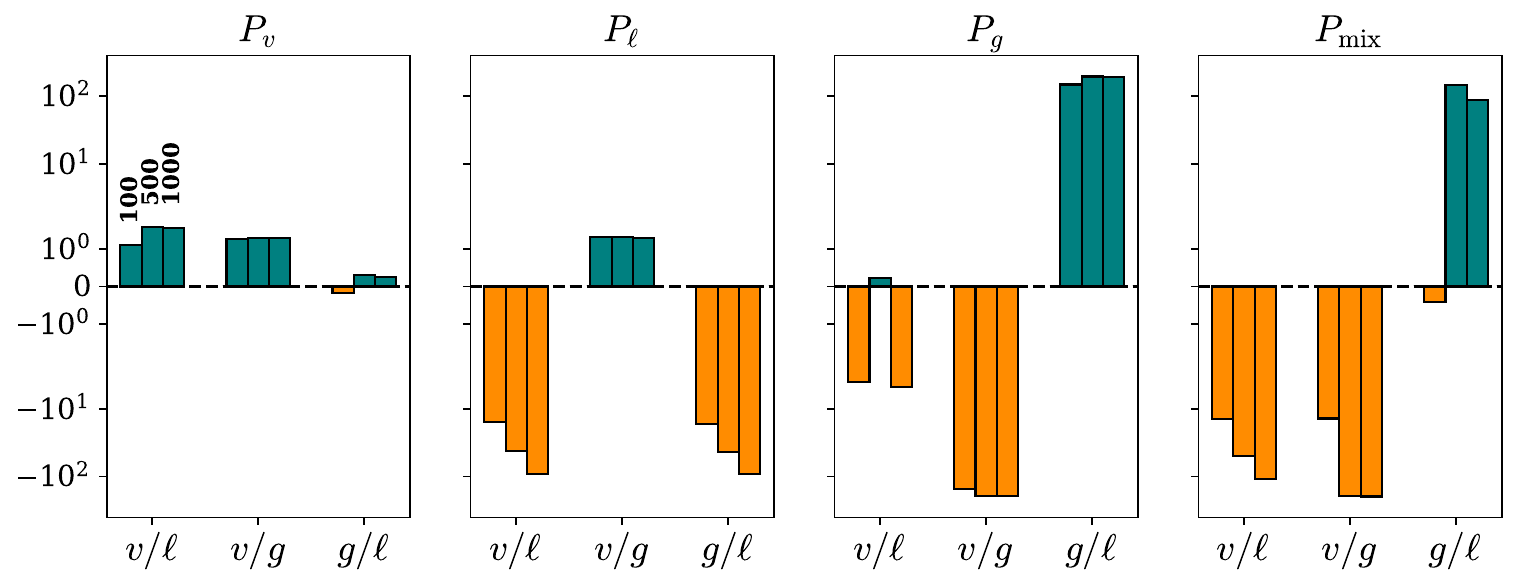}
    \caption{\justifying Bayes factors $\mathcal{B}^v_\ell, \mathcal{B}^v_g, $ and $\mathcal{B}^g_\ell$ in the considered populations $P_v$ (left panel), $P_\ell$ (left-center panel), $P_g$ (right-center panel), and $P_{\rm mix}$ (right panel). For brevity, we notate, e.g., $\log_{10}\mathcal{B}^v_\ell$ as $v/\ell$, which is to be understood as the preference for the hypothesis $\mathH_v$ over $\mathH_\ell$. The left, center, and right vertical bars in each panel represent the Bayes factors for the three different population sizes, $N_{\rm pop} = 100$, $500$, and $1000$, respectively. The horizontal dashed line in each panel represents a log Bayes factor of zero.} 
    \label{fig:Bayesfactorresultssinglepop}
\end{figure*}

With an SNR threshold of $\rho_{\rm thresh} = 20.0$, we find that only about $17\%$ of the sources are observable in all four populations, $P_v$, $P_\ell$, $P_g$, and $P_{\rm mix}$. We present the SNR distribution for the vacuum-GR population $P_v$ with $N_{\rm pop} = 1000$ sources in Fig.~\ref{fig:SNRdist}.

To infer the hyperposteriors under the local effect hypothesis, we draw $N = 5000$ samples of the vacuum-GR and local effect hyperparameters $\vlambda_v = \{K,\alpha,\beta\}$ and $\vlambda_\ell = \{f,(\mu_{A_\ell},\mu_{n_\ell}),(\sigma_{A_\ell},\sigma_{n_\ell})\}$ from narrow uniform priors around the true value, and weight them by the approximate hyperlikelihood in the local effect hypothesis, given by Eqs.~\eqref{eq:localhypersum},~\eqref{eq:localfirstpieceapprox}, and~\eqref{eq:localsecondpieceapprox}. The marginalized hyperposteriors of $f$ for all four populations $P_v$, $P_\ell$, $P_g$, and $P_{\rm mix}$ in the $N_{\rm pop} = 1000$ case are plotted in the top four panels of Fig.~\ref{fig:marginalhyperlike}. In $P_v$ and $P_\ell$, we find that the expectation $\langle f \rangle $ of $f$ (dashes vertical line) lies within $\sim 1\sigma$ of the true value $f^*$ (solid vertical line), indicating a robust recovery of the fractional population with migration torques, i.e. the local effect, in these populations. However, in $P_g$ and $P_{\rm mix}$, we recover $|\langle f\rangle - f^*| \approx 0.06$ representing biases $\sim 3.5\sigma$ and $\sim 1.5\sigma$ for the two populations, respectively. These systematic biases are a direct consequence of correlations between the migration torque parameters $A_\ell, n_\ell$ with the intrinsic strength of the global effect $\dot{G}~[\rm yr^{-1}]$. These correlations induce biases in the MLE estimates $\hat{A}_\ell, \hat{n}_\ell$ in all sources when analyzed under the local effect hypothesis. The extent to which these biases are measurable is reflected in the recovered value of $\langle f\rangle$. Similar results are obtained in the $N_{\rm pop} = 100, 500$ cases.

Similarly, for the global effect hypothesis, we obtain $5000$ samples from narrow uniform priors of $\vlambda_v$ and $\vlambda_g = \{\dot{G}\}$ around the true value and weight them by the corresponding hyperlikelihood~\eqref{eq:globallikelihoodapprox}. 
The marginalized hyperposteriors of $\dot{G}$ in the four populations for $N_{\rm pop} = 1000$ are plotted in the bottom panel of Fig.~\ref{fig:marginalhyperlike}. We find that the expectation $\langle\dot{G}\rangle$ is consistent at $\lesssim 0.01\sigma$ with the true value $\dot{G}^*$ in all four populations in the $N_{\rm pop} = 1000$ case. Notably, the width of the hyperlikelihood on $\dot{G}$ is broader by a factor of 2 in $P_\ell$ and $P_{\rm mix}$. This broadening is induced by the systematic biases in the MLE estimate of the source parameter $A_{g,i}$ due to non-null migration torques in these populations. Since these biases are independent, they average out over all observed sources, but still contribute to the measurement uncertainty of $\dot{G}$. This is also because the local effect is present in only $f = 0.5$ fraction of the sources, while all other sources have the correct MLE estimate, $A_{g,i} = \dot{G}^*$, such that the average over all observed sources is closer to the truth. Similar results are obtained in the $N_{\rm pop} = 100, 500$ cases.

We now calculate the Bayes factors $\mathcal{B}^v_\ell$ and $\mathcal{B}^v_g$ according to Eqs.~\eqref{eq:MCbayesfactorlocal} and~\eqref{eq:MCbayesfactorglobal} (and $\mathcal{B}^g_\ell = \mathcal{B}^v_\ell/\mathcal{B}^v_g$). We again draw $N = 5000$ samples of the vacuum-GR, local effect, and global effect hyperparameters, $\vlambda_v, \vlambda_\ell, \vlambda_g$, respectively, this time from their full priors (Table~\ref{tab:hyperpriorsall}), and weight them by the approximate hyperlikelihoods calculated in the appendices~\ref{app:vacuumhyperlikelihoodapp},~\ref{app:localhyperlikelihoodapp}, and~\ref{app:globalhyperlikelihoodapp} for the vacuum-GR, local effect, and global effect hypotheses, respectively. We plot the log Bayes factors for all four populations and the different population sizes in Fig.~\ref{fig:Bayesfactorresultssinglepop}. For brevity, we notate, e.g., $\log_{10}\mathcal{B}^v_\ell$ as $v/\ell$ in this plot. We find that the correct hypothesis is always preferred in all of the populations. Notably, in $P_g$ with $N_{\rm pop} = 100, 1000$, the local effect hypothesis is also preferred over vacuum-GR even though the true population does not have the local effect\footnote{The deviation from the trend in the $N_{\rm pop} = 500$ case may be attributed to one or more stochastic processes in the setup, which include generating the population, sampling from the hyperpriors, etc.}. This is again explained by the systematic biases induced on the MLE estimates of $A_\ell, n_\ell$ due to a non-null $\dot{G}$ in this population. The two hypotheses $\mathH_\ell, \mathH_g$ can still be disentangled by calculating $\mathcal{B}^g_\ell$, found in this case to be extremely favorable for the (correct) global effect hypothesis. This is represented by the ``$g/\ell$'' bars in the third panel. The effect of the size of the EMRI population is most evident in the calculation of $\mathcal{B}^g_\ell$ in the population $P_{\rm mix}$, where the preference for the global effect hypothesis over the local one flips when $N_{\rm pop}$ is small. This is because with fewer sources in the population, the measurement uncertainty on $\dot{G}$ is larger, and hence increasingly consistent with the null value.

\subsection{Robustness check}

To more broadly comment on the robustness of the hierarchical modeling framework for decoupling the considered beyond-vacuum-GR effects, i.e., migration torques and time-varying $G$, we now study the variability of the Bayes factor $\mathcal{B}^g_\ell$ as a function of the population's hyperparameters and size.

\begin{figure}[t]
    \centering
    \includegraphics[width=0.9\linewidth]{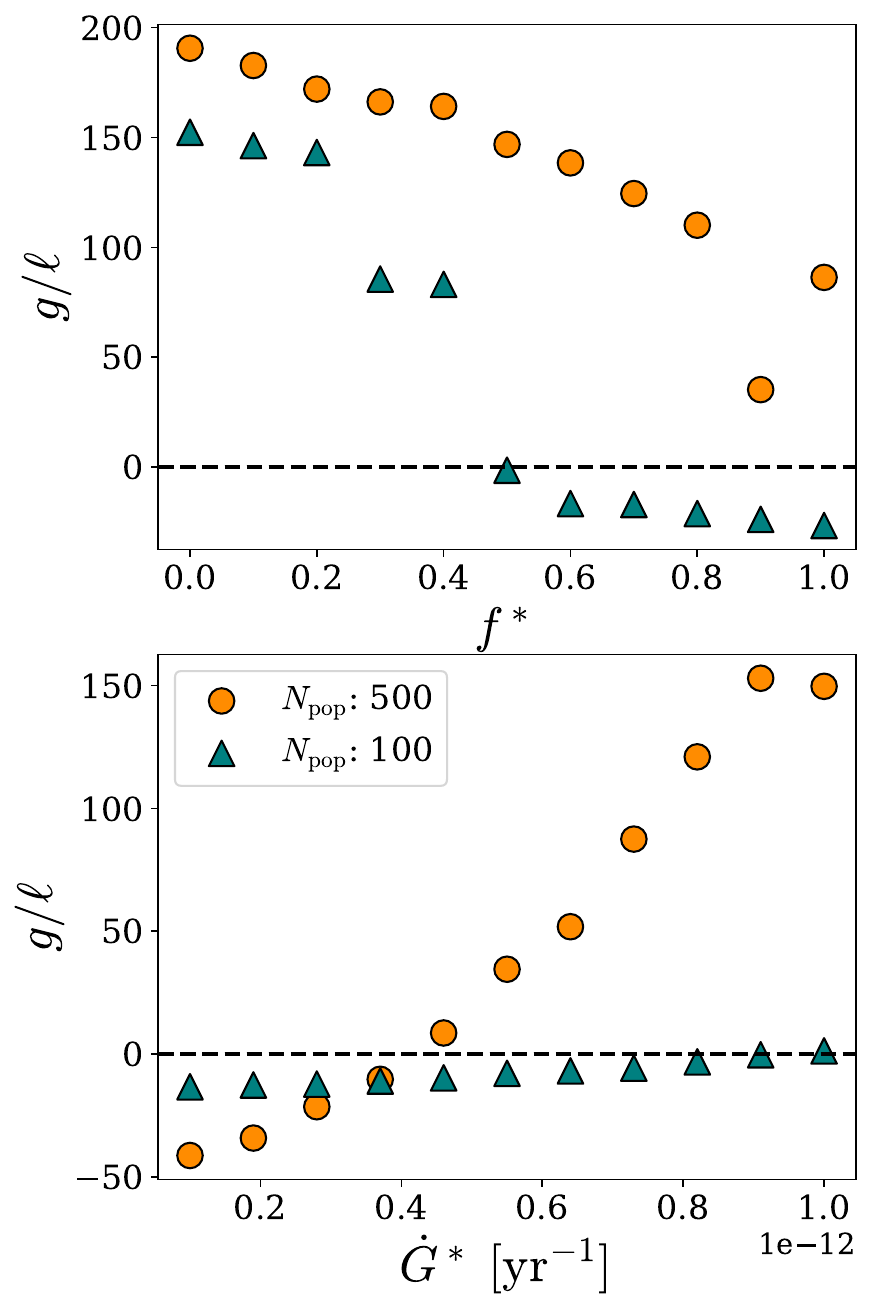}
    \caption{\justifying The log-Bayes factor $g/\ell := \log_{10}\mathcal{B}^g_\ell$ as a function of $f^*$ (top panel) and $\dot{G}^*$ (bottom panel) in $11$ different populations of size $N_{\rm pop} = 500$ (orange circles) and $N_{\rm pop} = 100$ (green triangles). In the top panels, $\dot{G}^* = 10^{-12} [\rm yr^{-1}]$ is fixed and $f^* \in [0.0,1.0]$. In the bottom panel, we fix $f^* = 0.5$ instead and $\dot{G}^* \in [10^{-13},10^{-12}]$. The horizontal line
in both panels represents $\log_{10}\mathcal{B}^g_\ell = 0.0$, i.e., for which the data does not favour either hypothesis.} 
    \label{fig:Bayesfactortrends}
\end{figure}

We consider 11 populations of fixed sizes $N_{\rm pop} = 100, 500$. In the first check, we fix $\dot{G}^* = 10^{-12} [\rm yr^{-1}]$ and vary $f^*$ linearly in the range $[0.0, 1.0]$. The setup otherwise is equivalent to that in the previous subsection. As shown in the top panels of Fig~\ref{fig:Bayesfactortrends}, the log-Bayes factor $\log_{10}\mathcal{B}^g_\ell$ scales inversely with $f^*$ for both population sizes, as expected. While the $N_{\rm pop} = 500$ population always prefers the global effect hypothesis, in the case of $N_{\rm pop} = 100$, the log-Bayes factor crosses zero at $f^* \approx 0.5$. This is because as $f^*$ increases in the population, the MLE estimate of the global effect's source parameter $\hat{A}_g$ is more biased across sources, and the inferred uncertainty in the hyperparameter $\dot{G}$ grows. This makes the hyperposterior on $\dot{G}$ more consistent with the null value, and hence the preference of the data for the global effect drops. In addition, we calculate the relative error $r_{\dot{G}} := |\dot{G}^* - \langle \dot{G} \rangle|/\dot{G}^*$ in all 11 populations. For $f \geq 0.8$, $r_{\dot{G}} \gtrsim 1.0$ for $N_{\rm pop} = 500$, i.e., the expectation value $\langle \dot{G} \rangle$ incurs significant biases with respect to the truth $\dot{G}^*$. This is due to the biases induced by the local effect across a larger fraction of the population. In the $N_{\rm pop} = 100$ case, the larger measurement uncertainties cover any systematic biases sufficiently well, such that there are no significant biases.

In the second check, we fix $f^* = 0.5$ and vary $\dot{G}^*$ linearly in the range $[10^{-13},10^{-12}]$ across the 11 populations. In populations of both sizes, $\log_{10}\mathcal{B}^g_\ell$ scales proportionally to the global effect hyperparameter $\dot{G}^*$ (bottom panel, Fig.~\ref{fig:Bayesfactortrends}). The log-Bayes factor $\log_{10}\mathcal{B}^g_\ell \lesssim 0.0$ for $N_{\rm pop} = 500$ ($=100$) when $\dot{G}^* \lesssim 4\times 10^{-13}~[\rm yr^{-1}]$ ($\lesssim 10^{-12}~[\rm yr^{-1}]$), i.e., the global effect hypothesis is no longer favored. This is because the data is less sensitive to the global effect if it is too weak, such that the measurement uncertainty in $\dot{G}$ grows.
In addition, at the zero crossing point of $\log_{10}\mathcal{B}^g_\ell$, even though both migration torques and time-varying $G$ are present in the population, they effectively ``cancel'' each other out. In such realizations of the population, the hierarchical framework will struggle to disentangle the two effects without supplementary checks. Finally, in the case of $N_{\rm pop} = 100$, we note that the log-Bayes factors are much closer to zero because of the larger uncertainty in measuring $\dot{G}$.

\subsection{A comment on other local effects}

As was shown in~\cite{Kejriwal:2023djc}, migration torques and time-varying $G$ can be severely degenerate in single source analyses across the EMRI parameter space. While other disk effects, such as accretion onto the MBH and dynamical friction~\cite{Kocsis:2011dr}, may also strongly correlate with time-varying $G$, it was shown in~\cite{Barausse:2014tra} that migration torques induce the largest correction to the orbital phase of the system. This is why migration torques are the main focus of this paper. Since each effect induces a distinct PN-order correction to the source's evolution, its measurability within the hierarchical framework may differ from that of migration torques and time-varying $G$ that we consider here, and should be studied in the future.

As another example, here we comment on the distinguishability of accretion (local effect) from time-varying $G$ (global effect) in the hierarchical framework. While accretion is not the dominant disk effect, it may be more severely degenerate with time-varying $G$ since both effects induce a correction to the fluxes at -4 PN order~\cite{Kocsis:2011dr, Barausse:2014tra}. We model the angular momentum flux as
\begin{align}
    \dot{L} = \dot{L}_{\rm GR}\left(1 + A'_\ell\left(\frac{p}{M}\right)^{n'_{\ell}} + A_g\left(\frac{p}{M}\right)^{n_{g}}\right)~\label{eq:accvstimevar}
\end{align}
where $A'_\ell := \dot{M}/M \lesssim 10^{-8}~[\rm yr^{-1}]$ is the accretion rate~\cite{Kocsis:2011dr} and $n'_\ell = 4$ corresponds to the -4 PN-order correction\footnote{We use the primed notation to distinguish accretion effect parameters from the migration torque effect parameters considered in the rest of the paper.}. We consider a population of size $N_{\rm pop} = 500$ sources, with $f^* = 0.2$ and $\dot{G}^* = 0.0$. $A'^*_\ell, n'^*_\ell$ are drawn from Eq.~\eqref{eq:priorlocal}, with $\vec{\mu}^*_\ell = (10^{-10},4)$ and $(\sigma^*_{A'_\ell},\sigma^*_{n'_\ell}) = (10^{-11},0.5)$.

In the analysis, we consider broad priors on $A'_\ell \in [0.0, 10^{-8}], n'_\ell \in [-10.0,10.0]$, and initially keep the same priors on $A_g$ as in Table~\ref{tab:priorsall}. The priors on the hyperparameters $f$ and $\dot{G}$ are also kept the same. Calculating the log-Bayes factors for the data's preference for the three hypotheses, we find $\log_{10}\mathcal{B}^v_\ell \approx -20.83$, $\log_{10}\mathcal{B}^v_g \approx 1.31$, and $\log_{10}\mathcal{B}^g_\ell \approx -22.25$, i.e., the correct (local effect) hypothesis is preferred over the other two (vacuum and global), even when the local and the global effects are of the same PN order.

As a final robustness check, we now extend the analysis prior on the global effect amplitude to $A_g \in [0.0, 10^{-8}]$, i.e., equal to the prior on $A'_\ell$. Given the mathematically identical form of the two effects (Eq.~\eqref{eq:accvstimevar}), these overlapping priors make the local and global effect parameters completely degenerate at the single source level. Recalculating the log-Bayes factors, we find that the local effect hypothesis is still preferred over the vacuum and the global ones, with $\log_{10}\mathcal{B}^v_\ell \approx -20.98$, $\log_{10}\mathcal{B}^v_g \approx 1.61$, and $\log_{10}\mathcal{B}^g_\ell \approx -22.59$. Although a more comprehensive analysis is beyond the scope of this work, this clearly demonstrates the hierarchical framework's robustness in disentangling the local and global effect hypotheses over a single-source analysis, where the two hypotheses would be indistinguishable.

\section{Discussion}\label{sec:discussion}

\subsection{Summary}

In this paper, we developed a hierarchical framework for GW populations with beyond-vacuum-GR effects, with the motivation to study how well local effects like astrophysical environments, which may be present in only a fraction of the population, can be decoupled from global ones like modified-GR contributions, which are expected to be present in all sources simultaneously. We consider specific models of local and global beyond-vacuum-GR effects: for the local effect, we choose migration torques induced by an accretion disk around the MBH. For the global effect, we model time-varying $G$, a modified gravity effect present universally with a constant amplitude $\dot{G} [\rm yr^{-1}]$. 

Considering four simulated EMRI populations, $P_v$, $P_\ell$, $P_g$, and $P_{\rm mix}$ and three different population sizes $N_{\rm pop} = 100, 500, 1000$, we find that the framework consistently recovers the fiducial true hypothesis that was used in constructing these populations, and even indicates whether both local and global effects are present in the population simultaneously. The recovered value of the global effect hyperparameter $\dot{G}$, i.e., the intrinsic strength of the time-varying $G$ effect, is consistent with the truth $\dot{G}^*$ at $\lesssim 0.01\sigma$ in all populations. However, the recovered value of the local effect hyperparameter $f$, i.e., the fractional population with migration torques, incurs systematic biases $\sim 3\sigma$ and $\sim 1.5\sigma$ compared to the true value $f^*$ in $P_g$ and $P_{\rm mix}$, respectively, for all three population sizes $N_{\rm pop} = 100, 500, 1000$. 

In addition, for two different populations of sizes $N_{\rm pop} = 100, 500$, we check that the Bayes factor $\mathcal{B}^g_\ell$ scales proportionally to the intrinsic strength $\dot{G}^* [\rm yr]^{-1}$ of the global effect in the population, and inversely with the fraction of sources affected by the local effect, $f^*$. We also find that the expectation of the global effect hyperparameter $\langle \dot{G} \rangle$ may significantly deviate from the ground truth if $f^* \geq 0.8$ when $N_{\rm pop} = 500$. For both cases $N_{\rm pop} = 100, 500$, we highlighted scenarios where the Bayes factor $\log_{10}\mathcal{B}^g_\ell \approx 0.0$ even when both effects are present in the population, such that the ability to tell them apart with the data is lost. Future studies should study such situations in more detail to further assess the robustness of the hierarchical framework.

\subsection{Future directions and outlook}

While we only consider circular-equatorial Kerr EMRIs in this paper, future studies should study the robustness of the hierarchical framework for beyond-vacuum-GR effects in more generic orbits, i.e., including the effects of eccentricity and inclination in the data. Such extensions may reduce parameter space degeneracies and, therefore, the biases incurred in recovering the local and global effect hyperparameters in mixed populations. More accurate models for EMRI waveforms and beyond-vacuum-GR effects are becoming increasingly available~\cite{Chapman-Bird:2025xtd,PhysRevLett.129.241103,Duque:2024mfw,Zi:2025lio,Duque:2025yfm,HegadeKR:2025dur,HegadeKR:2025rpr}. 

When constructing the populations in Sec.~\ref{sec:setupforgw}, we used independent prior pdfs on different classes of effects, and employed the same priors in the analysis setup. This simplification may not represent reality since the true distribution of the source parameters in the universe will not be known a priori. Additionally, both the local and global effect's strength may depend on the binary's vacuum-GR properties, such that the distributions are no longer independent. For more extensive applications of the hierarchical framework, future studies should investigate cases where all population parameters are realistically correlated and are drawn from distributions that differ from the priors used in the analysis.

A broader challenge for the hierarchical modeling framework lies in incorporating it within the proposed global fit pipeline for LISA data analysis~\cite{Vallisneri:2008ye,Littenberg:2023xpl,Katz:2024oqg}. The global fit aims to infer all source types in the data stream simultaneously, and will be expensive to evaluate even in the simplest vacuum-GR hypothesis $\mathH_v$. Redoing the global fit for the alternative hypotheses $\mathH_\ell$ and $\mathH_g$ (which also include additional parameters) may not be practically feasible. Yet, it may be possible to perform hypothesis tests in a post-processing style analysis by bias-correcting and importance-weighting the samples obtained in the $\mathH_v$ analysis~\cite{Kejriwal:2025upp}. This procedure effectively provides a new set of samples from the hyperposterior in the alternative hypothesis and is particularly well-suited for beyond-vacuum-GR type effects (see~\cite{Kejriwal:2025upp} for more details). Recently, simulation-based inference (SBI) techniques have been proposed for feasible hierarchical modeling of GW sources~\cite{Srinivasan:2025etu}, and may enable efficient modeling and inference of beyond-vacuum-GR effects using source populations. 

Finally, we caution that some systematic biases may persist in the hierarchical framework, e.g., due to inaccuracies in waveform modeling which can mimic global deviations from the null hypothesis, as discussed in~\cite{Moore:2021eok}. In our framework, these systematic biases can be understood to be induced by modeling errors, such as missing higher-order corrections in the vacuum-GR evolution~\cite{Burke:2023lno} or insufficiently accurate template waveforms~\cite{Khalvati:2025znb}. Because such waveform systematics will be pervasive throughout the observed population, they may strongly bias the inference of a global effect such as the intrinsic strength $\dot{G}$, even within the hierarchical framework. Our setup can be readily employed to study such systematics in the future. Extensions of the current framework, shaped by these careful considerations, will pave the way for robust and realistic tests of GR in the presence of astrophysical environments with the next generation of GW detectors.

\acknowledgments
SK thanks Patrick Meyers and Michele Vallisneri for useful discussions during a research visit to ETH Zurich. SK also acknowledges the computing resources provided by the NUS IT Research Computing group and the support of the NUS Research Scholarship.
EB acknowledges support from the European Union’s Horizon ERC Synergy Grant ``Making Sense of the Unexpected in the Gravitational-Wave Sky'' (Grant No. GWSky-101167314) and the PRIN 2022 grant ``GUVIRP - Gravity tests in the UltraViolet and InfraRed with Pulsar timing''.
This work has been supported by the Agenzia Spaziale Italiana (ASI), Project n. 2024-36-HH.0, ``Attività per la fase B2/C della missione LISA''.  

\bibliography{references}

\appendix

\section{Selection effects for beyond-vacuum-GR hyperparameters}~\label{app:constmargselfun}

We denote the null vacuum-GR hypothesis as $\mathH_{v}$ and the alternative local and global effect hypotheses jointly as $\mathH_a = \mathH_\ell, \mathH_g$. We decompose the hyperparameter set as $\vlambda = \{\vlambda_{v},\vlambda_{a}\}$, and the source parameters $\vtheta = \{\vtheta_{v},\vtheta_{a}\}$, where a subscript $_v$ denotes vacuum-GR and $_a$ denotes beyond-vacuum-GR effect parameters. In this notation, the selection function can be written as~\cite{Mandel_2019}
\begin{align}
    \alpha(\vlambda|\mathH_{a}) = \int{\rm d}\vtheta~ p_{\rm det}(\vtheta)\pi(\vtheta|\vlambda,\mathH_{a})~\label{appeq:selectionfunctionsuperset}
\end{align}
where $p_{\rm det}(\vtheta)$ is the probability of detecting a source with parameters $\vtheta$, and $\pi(\vtheta|\vlambda,\mathH_a)$ is the prior pdf of $\vtheta$ for the given hyperparameter set $\vlambda$ in the alternative hypothesis.

Since $\mathH_a$ describes a set of (perturbative) beyond-vacuum-GR effects, we argue that the detection probability is approximately independent of $\vtheta_{a}$, such that
\begin{align}
    p_{\rm det}(\vtheta) \approx p_{\rm det}(\vtheta_{v}).
\end{align}
Additionally, if we assume that the priors of $\vtheta_v$ and $\vtheta_a$ are independent, i.e., $\pi(\vtheta|\vlambda,\mathH_a) = \pi(\vtheta_v|\vlambda_v)\pi(\vtheta_a|\vlambda_a)$, we get
\begin{align}
    \alpha(\vlambda_{a}|\mathH_{a}) \approx \mathcal{C}_{v}\int{\rm d}\vtheta_{a}~  p(\vtheta_{a}|\vlambda_{a}).~\label{appeq:margselfun}
\end{align}
Here, $\mathcal{C}_{v} := \int {\rm d}\vlambda_{v}{\rm d}\vtheta_{v}~ p_{\rm det}(\vtheta_{v})\pi(\vtheta_{v}|\vlambda_{v})$
is a constant. Since the remaining integral in Eq.~\eqref{appeq:margselfun} is independent of the detection probability, it is effectively calculated over the full space of $\vtheta_{a}$. In this paper, we only consider proper priors for the beyond-vacuum-GR effects, and hence $\int {\rm d}\vtheta_a~\pi(\vtheta_a|\vlambda_a) = 1$ for arbitrary $\vlambda_a$. Therefore, the selection function $\alpha(\vlambda_a|\mathH_a) \approx \mathcal{C}_v$ is constant.

\section{Derivations of the approximate hyperlikelihoods}

In this appendix, we explicitly derive the analytical approximations for the hyperlikelihoods $p(\htheta_i|\vlambda,\mathH)$~\eqref{eq:sourceintegral} in the vacuum-GR, local effect, and global effect hypotheses. We validate our construction against the direct Monte Carlo integration of Eq.~\eqref{eq:sourceintegral} in the final subsection.

\subsection{Linear-bias approximation}\label{app:linearbiasapprox}

For notational convenience, we introduce an alternate decomposition of the inferred source parameter set $\vtheta := \{\vpsi,\vvarphi\}$. Here, $\vpsi$ is the set of parameters that are inferred in the analysis, and $\vvarphi$ are the remaining model parameters fixed to some null value $\vvarphi_0$ in the analysis. We re-emphasize that $\vvarphi_0$ is different from the true value of the uninferred parameters $\vvarphi^*$ in general. Then, from the linear-bias approximation, the MLE estimate $\hpsi_i$ of the inferred parameters is~\cite{Kejriwal:2023djc}
\begin{align}
    \hpsi_i \approx \vpsi_i^* + (\Gamma_i^{\psi\psi})^{-1}\cdot\Gamma_i^{\psi\varphi}\cdot(\vvarphi_i^* - \vvarphi_{i,0})\label{eq:CVnested}
\end{align}
where $\Gamma_i$ is the Fisher information matrix (FIM) of the $i^{\rm th}$ observed source. It is calculated at $\vtheta_i^*$. $\Gamma_i^{\psi\psi}$, and $\Gamma_i^{\psi\varphi}$ are submatrices of $\Gamma_i$ corresponding to the $\vpsi,\vpsi$ and $\vpsi,\vvarphi$ elements, respectively. 

\subsection{Vacuum-GR}\label{app:vacuumhyperlikelihoodapp}

\begin{figure}
    \centering
    \includegraphics[width=0.8\linewidth]{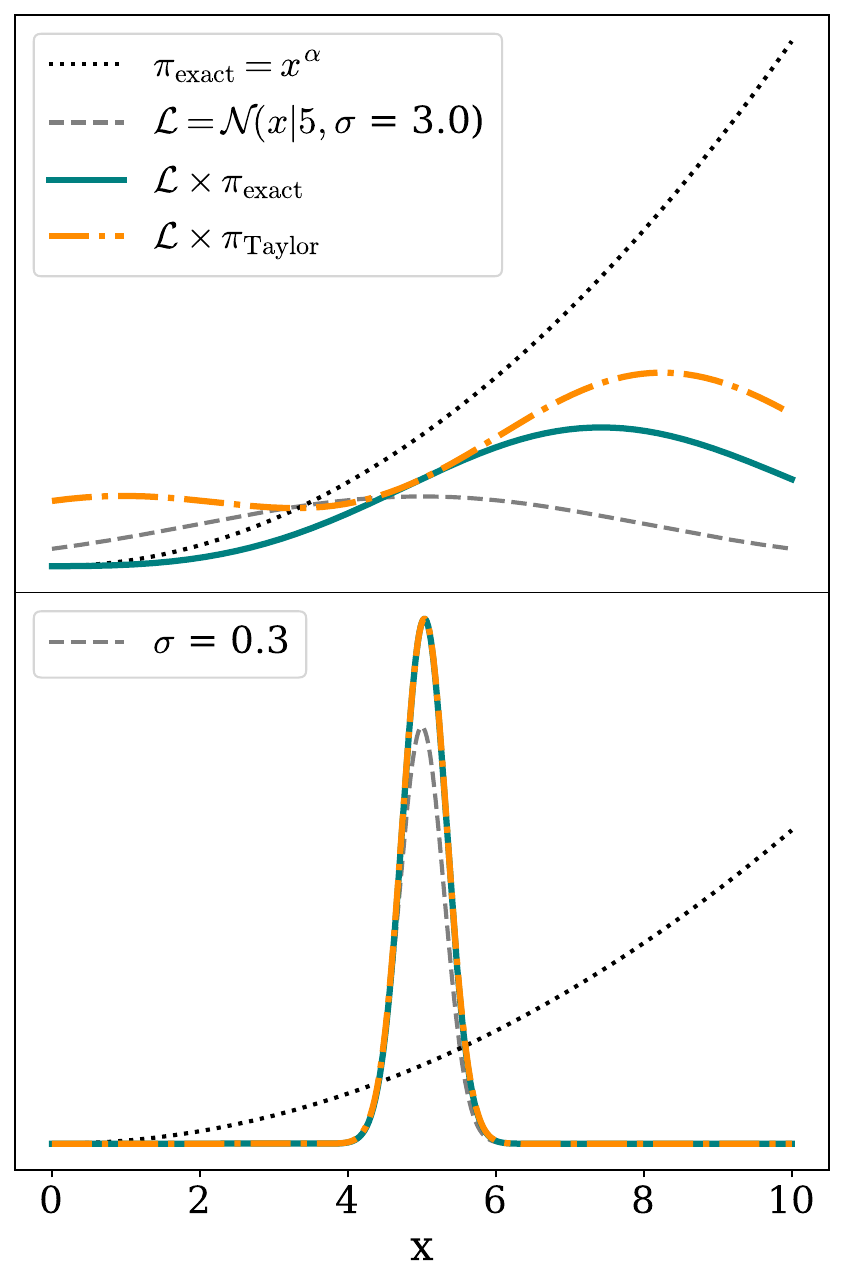}
    \caption{\justifying{1-dimensional visualization of the product of a normal likelihood $\mathcal{L}(x) = \mathcal{N}(x|5,\sigma)$ with a power-law prior on $x$, $p(x) = x^\alpha$ where $\alpha = 2$ is the slope hyperparameter. In both panels, the green solid curve represents this product in its exact form, while the orange dash-dotted curve represents the approximate product, where $p(x)$ is Taylor-expanded to the second order in $x - 5$ about $5$. In the top panel, the normal likelihood has a large standard deviation $\sigma = 3.0$, such that its rate of change is similar to that of the power-law prior, and the Taylor approximation fails. When we set $\sigma = 0.3$ as in the bottom panel, the normal likelihood varies at much smaller scales (as expected for GW data analysis), and the Taylor approximation holds well.}}
    \label{fig:taylortoy}
\end{figure}

In the vacuum-GR hypothesis, $\mathH_v$, the inferred set of parameters is $\vpsi \equiv \{\vec{v}\}$, while the fixed set is $\vvarphi \equiv \{\vec{\ell},\vec{g}\}$, which assume the null value $\vvarphi_{i,0} = \{\vec{0}, \vec{0}\}$ in the analysis. Correspondingly, the MLE $\hat{\psi}_i \equiv \hat{v}_i$ for the $i^{\rm th}$ source is obtained from Eq.~\eqref{eq:CVnested} and the source likelihood~\eqref{eq:normallikelihood} can be written as
\begin{align}
    p(\vec{d}_i|\vec{v},\mathH_v) = \mathcal{N}(\vec{v}|\hat{v}_i, \left(\Gamma_i^{vv}\right)^{-1}).
\end{align}
The prior pdf in the vacuum-GR hypothesis, $\pi(\vec{v}|\vlambda_v)$ is given in Eq.~\eqref{eq:priorvacuum}. Substituting in Eq.~\eqref{eq:sourceintegral},
\begin{align}
    p(\vec{d}_i|\vlambda_v,\mathH_v) = \int {\rm d}\vec{v} ~\mathcal{N}(\vec{v}|\hat{v}_i,\left(\Gamma_i^{vv}\right)^{-1})\pi(\vec{v}|\vlambda_v).~\label{eq:vacLSAfull}
\end{align}
While this integral is over the full space of $\vec{v}$, the integrand is a product of a normal distribution centered at $\hat{v}_i$ and a power-law in $\vec{v}$. Thus, at large distances $\Delta\vec{v}_i := \vnorm{\vec{v}-\hat{v}_i} \gg \hat{v}_i$ away from $\hat{v}_i$, the integrand is exponentially suppressed, such that 
\begin{align}
    \mathcal{N}(\vec{v}|\hat{v}_i,\left(\Gamma_i^{vv}\right)^{-1})\pi(\vec{v}|\vlambda_v) \approx \mathcal{N}(\vec{v}|\hat{v}_i,\left(\Gamma_i^{vv}\right)^{-1})\pi(\vec{v}|\vlambda_v)_{\rm Taylor}~\label{eq:taylorintegrand}
\end{align}
where
\begin{align*}
    \pi(\vec{v}|\vlambda_v)_{\rm Taylor} := ~& \pi(\hat{v}_i|\vlambda_v) + \Delta\vec{v}_i^k\partial_k\pi(\hat{v}_i|\vlambda_v) \\
    &+ \Delta\vec{v}_i^k\Delta\vec{v}_i^m\frac{\partial_k\partial_m\pi(\hat{v}_i|\vlambda_v)}{2}\numberthis
\end{align*}
is the Taylor expansion of the power-law prior pdf to quadratic order in $\Delta\vec{v}_i$. Here, $\partial_k\pi(\hat{v}_i|\vlambda_v):= \partial \pi(\vec{v}|\vlambda_v)/\partial\vec{v}^k|_{\hat{v}_i}$ and we assume the Einstein summation convention over the $k$ and $m$ indices.

We remark that the order in $\Delta\vec{v}_i$ to which the prior pdf should be Taylor expanded, such that Eq.~\eqref{eq:taylorintegrand} holds, is determined by the scale at which the exponential suppression dominates the integrand around $\hat{v}_i$. This subsequently relies on the relative rate of change of the power-law and the normal distributions. We demonstrate this scale-dependence with a simple one-dimensional example in Fig.~\ref{fig:taylortoy}. Since we are working in the LSA, the normal approximated likelihoods are tightly centered around $\hat{v}_i$ and the power-law prior evolves at much slower rates, such that a quadratic-order Taylor expansion suffices. We also verified that Eq.~\eqref{eq:taylorintegrand} holds for typical sources in the results presented in Sec.~\ref{sec:results}.

Substituting Eq.~\eqref{eq:taylorintegrand} in~\eqref{eq:vacLSAfull}, we get
\begin{align}
    ~p(\hat{v}_i|\vlambda_v,\mathH_v) \approx \pi(\hat{v}_i|\vlambda_v) + \left(\left(\Gamma_i^{vv}\right)^{-1}\right)^{km}\frac{\partial_k\partial_m\pi(\hat{v}_i|\vlambda_v)}{2}~\label{eq:vachyperlikelihoodapprox}
\end{align}
using standard results of multinormal integration over the full space,
\begin{align*}
    \int{\rm d}\vec{v}~\mathcal{N}(\vec{v}|\hat{v}_i,\left(\Gamma_i^{vv}\right)^{-1}) &= 1,\numberthis\\
    \int {\rm d}\vec{v}~\mathcal{N}(\vec{v}|\hat{v}_i,\left(\Gamma_i^{vv}\right)^{-1})\Delta\vec{v}^k_i = \mathbb{E}[\Delta\vec{v}_i^k] &= 0,\numberthis\\
    \int {\rm d}\vec{v}~\mathcal{N}(\vec{v}|\hat{v}_i,\left(\Gamma_i^{vv}\right)^{-1})\Delta\vec{v}^k_i\Delta\vec{v}^m_i &= \mathbb{E}[\Delta\vec{v}_i^k\Delta\vec{v}_i^m] \\
    &= \left(\left(\Gamma_i^{vv}\right)^{-1}\right)^{km}\numberthis
\end{align*}
for all $k,m$. Here the expectation $\mathbb{E}[\cdot]$ is calculated over the LSA likelihood $\mathcal{N}(\vec{v}|\hat{v}_i,(\Gamma_i^{vv})^{-1})$.

\subsection{Local effect}~\label{app:localhyperlikelihoodapp}

In the local effect hypothesis, $\mathH_\ell$, the inferred set of parameters is $\vpsi \equiv \{\vec{v},\vec{\ell}\}$, while the uninferred set is $\vvarphi \equiv \{\vec{g}\}$. Hence, with the MLE $\hpsi_i$ given by Eq.~\eqref{eq:CVnested} and the prior $\pi(\vtheta|\vlambda,\mathH_\ell) \equiv \pi(\vec{v}|\vlambda_v)\pi(\vec{\ell}|\vlambda_\ell)$ as described in Eqs.~\eqref{eq:priorvacuum},~\eqref{eq:priorlocal}, the hyperlikelihood can be written in two pieces
\begin{align}
    p(\hpsi_i|\vlambda,\mathH_\ell) = p^{(1)}(\hpsi_i|\vlambda,\mathH_\ell) + p^{(2)}(\hpsi_i|\vlambda,\mathH_\ell)~\label{eq:localhypersum}
\end{align}
where
\begin{align*}
    p^{(1)}(\hpsi_i|\vlambda,\mathH_\ell) := (1 - f) \int {\rm d}\vpsi~&\mathcal{N}(\vpsi|\hpsi_i,(\Gamma_i^{\psi\psi})^{-1})\times\\
    &\delta^2(\vec{\ell})\pi(\vec{v}|\vlambda_v),\numberthis \label{appeq:localintegralone}
\end{align*}
and
\begin{align*}
    p^{(2)}(\hpsi_i|\vlambda,\mathH_\ell) := f \int{\rm d}\vpsi ~&\mathcal{N}(\vpsi|\hpsi_i,(\Gamma_i^{\psi\psi})^{-1})\times\\
    &\mathcal{N}(\vec{\ell}|\vec{\mu}_\ell,\Sigma_\ell)\pi(\vec{v}|\vlambda_v)\numberthis\label{appeq:localintegraltwo}
\end{align*}
can be evaluated separately. In the following, we notate the dimension of a generic vector $\vec{a}$ as $D_a$.

\subsubsection{Approximating the first piece}
In approximating~\eqref{appeq:localintegralone}, we first evaluate the integral over $\vec{\ell}$, such that the product of the normal distribution and the Dirac-delta prior gives us a joint normal pdf with $\vec{\ell} = \vec{0}$ fixed,
\begin{align}
    p^{(1)}(\hpsi_i|\vlambda,\mathH_\ell) = (1-f)\int{\rm d}\vec{v}~\mathcal{N}(\{\vec{v},\vec{0}\}|\hpsi_i,(\Gamma_i^{\psi\psi})^{-1})\pi(\vec{v}|\vlambda_v).~\label{eq:localonestart}
\end{align}
The joint normal pdf can be decomposed into a conditional and a marginal~(see, e.g., section 2.3 of~\cite{bishop2006pattern}),
\begin{align}
    \mathcal{N}(\{\vec{v},\vec{0}\}|\hpsi_i,(\Gamma_i^{\psi\psi})^{-1}) = \mathcal{N}(\vec{v}|v_i^\dagger,(\Gamma_i^{vv})^{-1})\times\mathcal{S}^{(1)}_i
\end{align}
where,
\begin{align}
     v^\dagger_i &:= \hat{v}_i + (\Gamma_i^{vv})^{-1}\cdot\Gamma_i^{v\ell}\cdot\hat{\ell}_i.\label{eq:localonevdagger}
\end{align}
is the mean of the normal pdf.~\footnote{Note that the form of $v_i^\dagger$ is objectively equivalent to the linear-bias formula for nested models~\eqref{eq:CVnested} with the ``true'' source parameter set $\vpsi_i^*:=\{\hat{v}_i,\hat{\ell}_i\}$ and where $\vvarphi :=\vec{\ell}=\vec{0}$ in the ``null'' hypothesis. This is directly related to the assumption of the LSA in obtaining~\eqref{eq:CVnested}, such that the MLE in the null hypothesis is precisely the expectation of the (normal) LSA likelihood conditioned at $\vec{\ell} = \vec{0}$.} $\mathcal{N}(\vec{v}|v_i^\dagger,(\Gamma_i^{vv})^{-1})$ conditioned at $\vec{\ell}=\vec{0}$, and
\begin{widetext}
    \begin{align}
        \mathcal{S}_i^{(1)} := \frac{(2\pi)^{D_v/2}}{|\Gamma_i^{vv}|^{1/2}}\frac{|\Gamma_i^{\psi\psi}|^{1/2}}{(2\pi)^{D_\psi/2}}\exp\left[-\frac{1}{2}\hat{\ell}_i^T\cdot(\Gamma_i^{\ell\ell}-\Gamma_i^{\ell v}\cdot(\Gamma_i^{vv})^{-1}\cdot\Gamma_i^{v\ell})\cdot\hat{\ell}_i\right]`\label{eq:localonenormalization}
    \end{align}
\end{widetext}
is the normalized marginal probability of $\vec{\ell}=0$ given the data realization $\hat{\ell}_i$. Substituting in~\eqref{eq:localonestart}, 
\begin{align*}
    p^{(1)}(\hpsi_i|\vlambda,\mathH_\ell) = &(1-f)\mathcal{S}_i^{(1)}\times\\
    &\int{\rm d}\vec{v}~\mathcal{N}(\vec{v}|v_i^\dagger,(\Gamma_i^{vv})^{-1})\pi(\vec{v}|\vlambda_v).\numberthis
\end{align*}
Thus, we obtain an integral similar in form to Eq.~\eqref{eq:vacLSAfull}, where the integrand is the product of a normal distribution centered at $v_i^\dagger$ and a power-law that evolves at a much slower rate. Following a similar argument as for the vacuum-GR integral (Sec.~\ref{app:vacuumhyperlikelihoodapp}), the power-law prior can be Taylor expanded to second order in $\vnorm{\vec{v}-v_i^\dagger}$ about $v_i^\dagger$ to obtain the final result for the first piece of the local effect hyperlikelihood,
\begin{align*}
    p^{(1)}(\hpsi_i|\vlambda,\mathH_\ell) \approx &(1-f)~\mathcal{S}_i^{(1)}\times\\
    &\left[\pi(v_i^\dagger|\vlambda_v) + ((\Gamma^{vv}_i)^{-1})^{km}\frac{\partial_k\partial_m \pi(v_i^\dagger|\vlambda_v)}{2}\right].\numberthis\label{eq:localfirstpieceapprox}
\end{align*}

\subsubsection{Approximating the second piece}\label{app:secondpiece}
To approximate the second piece~\eqref{appeq:localintegraltwo}, we first note that the integrand has a product of (different dimensional) Gaussian pdfs. We evaluate this product by defining a projection matrix $P$ of dimensionality $D_\ell \times D_\psi$ such that the local effect parameter vector can be written as $\vec{\ell} = P\vpsi$. Then~\cite{Hogg:2020jwh},
\begin{align}
    \mathcal{N}(\vpsi|\hpsi_i,(\Gamma_i^{\psi\psi})^{-1})\mathcal{N}(\vec{\ell}|\vec{\mu}_\ell,\Sigma_\ell) = \mathcal{S}_i^{(2)}\mathcal{N}(\vpsi|\tilde{\psi}_i,\tilde\Gamma_i^{-1})
\end{align}
where
\begin{align}
    \tilde{\Gamma}_i &:= \Gamma_i^{\psi\psi} + P^T\Sigma_\ell^{-1}P, \label{eq:localtwofisher}\\
    \tilde{\psi}_i \equiv \{\tilde{v}_i,\tilde{l}_i\} &:= \tilde{\Gamma}_i^{-1}(\Gamma_i^{\psi\psi}\hpsi_i + P^T\Sigma_\ell^{-1}\vec{\mu}_\ell), \label{eq:localtwotilde}\\
    \mathcal{S}_i^{(2)} &:= \mathcal{N}(\vec{\mu}_\ell|\hat{\ell}_i, P(\Gamma_i^{\psi\psi})^{-1}P^T + \Sigma_\ell). \label{eq:localtwonormalization}
\end{align}
Substituting in~\eqref{appeq:localintegraltwo},
\begin{align}
    p^{(2)}(\hpsi_i|\vlambda,\mathH_\ell) &= f~\mathcal{S}_i^{(2)}\int{\rm d}\vpsi~\mathcal{N}(\vpsi|\tilde{\psi}_i,\tilde\Gamma_i^{-1})\pi(\vec{v}|\vlambda_v)\\
    &=f~\mathcal{S}_i^{(2)}\int{\rm d}\vec{v}~\mathcal{N}(\vec{v}|\tilde{v}_i,(\tilde\Gamma_i^{-1})^{vv})\pi(\vec{v}|\vlambda_v).
\end{align}
In the second equality, we evaluate the integral over $\vec{\ell}$ to obtain the marginal likelihood of $\vec{v}$.
Taylor-expanding $\pi(\vec{v}|\vlambda_v)$ about $\tilde{v}_i$ to second order in $\vnorm{\vec{v}-\tilde{v}_i}$, we obtain the final form of the second integral piece, 
\begin{align*}
    p^{(2)}(\hpsi_i|\vlambda,\mathH_\ell)\approx &f~ \mathcal{S}^{(2)}_i \times \\
    &\left[\pi(\tilde{v}_i|\vlambda_v) + \left(\tilde{\Gamma}_i^{-1}\right)^{km}\frac{\partial_k\partial_m \pi(\tilde{v}_i|\vlambda_v)}{2}\right],\numberthis~\label{eq:localsecondpieceapprox}
\end{align*}
where we use for brevity the equivalence $(\tilde\Gamma_i^{-1})^{km} \equiv ((\tilde\Gamma_i^{-1})^{vv})^{km}$ for indices $k,m$ that cycle only through the vacuum-GR parameters.

\subsection{Global effect}~\label{app:globalhyperlikelihoodapp}

\begin{figure*}
    \centering
    \includegraphics[width=0.9\linewidth]{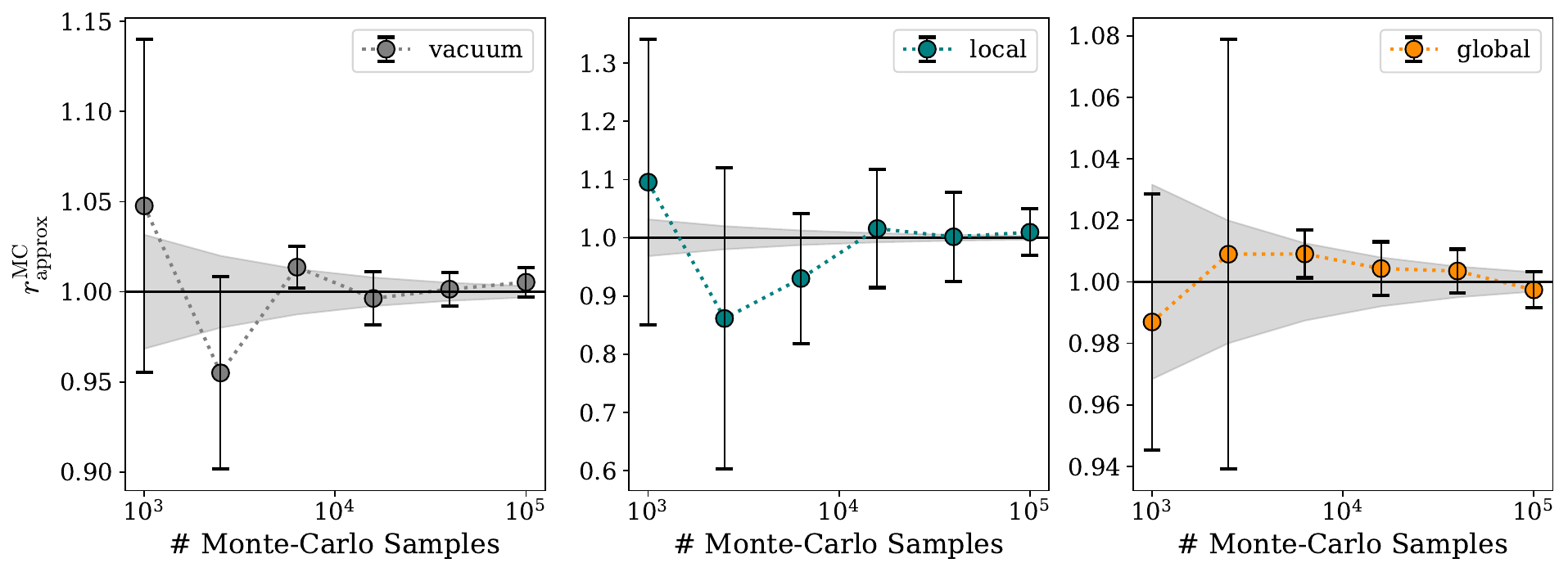}
    \caption{\justifying{Validating the approximate hyperlikelihoods calculated in the vacuum-GR hypothesis (Eq.~\eqref{eq:vachyperlikelihoodapprox}, left panel), the local effect hypothesis (Eqs.~\eqref{eq:localhypersum},~\eqref{eq:localfirstpieceapprox}, and~\eqref{eq:localsecondpieceapprox}, middle panel), and the global effect hypothesis (Eq.~\eqref{eq:globallikelihoodapprox}, right panel) against Monte-Carlo integral of the exact expression (Eq.~\eqref{eq:sourceintegral}) for EMRIs. The vertical axis plots the ratio of the Monte Carlo integral to the approximate hyperlikelihood, $r^{\rm MC}_{\rm approx}$. The horizontal black solid line in each panel represents $r^{\rm MC}_{\rm approx} = 1.0$, i.e., where the two approaches match exactly. The vertical bars represent $1\sigma$ errors on the Monte Carlo integral for a given sample size, estimated empirically from five independent trials, while the shaded region represents the ``theoretical'' 1$\sigma$ error in the Monte Carlo integral approximation with $N$ samples, $1/\sqrt{N}$ .}}
    \label{fig:integralvalidation}
\end{figure*}

Finally, in the global effect hypothesis, $\mathH_g$, the inferred set of parameters is $\vpsi = \{\vec{v},\vec{g}\} = \{\vec{v},A_g\}$ and $\vvarphi=\{\vec{\ell}\}$ is the uninferred set. Given the MLE $\hpsi_i = \{\hat{v}_i,\hat{g}_i\} = \{\hat{v}_i,\hat{A}_{g,i}\}$ from Eq.~\eqref{eq:CVnested}, and the prior $\pi(\vtheta|\vlambda,\mathH_g) \equiv \pi(\vec{v}|\vlambda_v)\pi(\vec{g}|\vlambda_g)$ from Eqs.~\eqref{eq:priorvacuum},~\eqref{eq:priorglobal}, the hyperlikelihood can be written as
\begin{align*}
    p(\hpsi_i|\vlambda,\mathH_g) = \int {\rm d}\vpsi~&\mathcal{N}(\vpsi|\hpsi_i,(\Gamma_i^{\psi\psi})^{-1})\times\\
    &\delta(A_g - \dot{G})\pi(\vec{v}|\vlambda_v).\numberthis \label{eq:globalhyperlikefull}
\end{align*}
It is notably similar in form to the first piece of the hyperlikelihood in the local effect case~\eqref{appeq:localintegralone}, and can be solved following a similar argument.

We first evaluate the integral over $\vec{g} \equiv \{A_g\}$. We have
\begin{align}
    p(\vpsi_i|\vlambda,\mathH_g) = \int {\rm d}\vec{v}~\mathcal{N}(\{\vec{v},\dot{G}\}|\hpsi_i,(\Gamma_i^{\psi\psi})^{-1})\pi(\vec{v}|\vlambda_v).~\label{appeq:globalmarginalizedoverA}
\end{align}
Then, we decompose the joint normal pdf into a conditional and a marginal~\cite{bishop2006pattern},
\begin{align}
    \mathcal{N}(\{\vec{v},\dot{G}\}|\hpsi_i,(\Gamma_i^{\psi\psi})^{-1}) = \mathcal{N}(\vec{v}|v_i^\ddagger,(\Gamma_i^{vv})^{-1})\times\mathcal{S}_i~\label{appeq:globalintegrand}
\end{align}
where,
\begin{align}
    v_i^\ddagger := \hat{v}_i + (\Gamma^{vv}_i)^{-1}\cdot\Gamma_i^{vg}\cdot(\hat{A}_{g,i}-\dot{G})\label{eq:globalmean}
\end{align}
and
\begin{widetext}
\begin{align}
    \mathcal{S}_i := \frac{(2\pi)^{D_v/2}}{|\Gamma_i^{vv}|^{1/2}}\frac{|\Gamma_i^{\psi\psi}|^{1/2}}{(2\pi)^{D_\psi/2}}\exp\left[ -\frac{1}{2}(\Gamma_i^{gg}-\Gamma_i^{gv}\cdot(\Gamma_i^{vv})^{-1}\cdot\Gamma_i^{vg})(\hat{A}_{g,i}-\dot{G})^2 \right].\label{eq:globalnormalization}
\end{align}
\end{widetext}
Substituting Eq.~\eqref{appeq:globalintegrand} into~\eqref{appeq:globalmarginalizedoverA}, and following arguments similar to Appendix~\ref{app:secondpiece}, the analytical approximation of the global effect hyperlikelihood is given as
\begin{align*}
    p(\hpsi_i|\vlambda,\mathH_g) \approx \mathcal{S}_i\Bigg[&\pi(v_i^\ddagger|\vlambda_v) + \\
    &((\Gamma_i^{vv})^{-1})^{km}\frac{\partial_k\partial_m \pi(v_i^\ddagger|\vlambda_v)}{2} \Bigg]\numberthis.~\label{eq:globallikelihoodapprox}
\end{align*}

\subsection{Validating the approximate hyperlikelihoods}~\label{app:validation}

To summarize, in this section, we have analytically approximated the hyperlikelihood~\eqref{eq:sourceintegral} by assuming the validity of the LSA and utilizing various standard properties of normal distributions. Our approximate results are given in the vacuum-GR hypothesis, $\mathH_v$, by Eq.~\eqref{eq:vachyperlikelihoodapprox}; for the local-effect hypothesis, $\mathH_\ell$, by Eqs.~\eqref{eq:localhypersum},~\eqref{eq:localfirstpieceapprox}, and~\eqref{eq:localsecondpieceapprox}; and in the global-effect hypothesis, $\mathH_g$, by Eq.~\eqref{eq:globallikelihoodapprox}. 

We will now validate the approximate results in the EMRI context by comparing them with the direct Monte Carlo integral of Eq.~\eqref{eq:sourceintegral}. In each hypothesis, we initialize an EMRI source on circular and equatorial orbits evolving according to Eq.~\eqref{eq:angularmomentumbvgr}, with true source parameters $\vtheta^*$ and population hyperparameters $\vlambda^*$ satisfying the respective hypothesis: the vacuum-GR parameters are fixed across all three sources and are chosen to be representative of typical EMRIs, while the local and global effect parameters are non-null only in the local effect and global effect hypothesis, respectively. By construction, since each EMRI is initialized in its true hypothesis, the MLE $\hpsi$ of the inferred parameters is not biased, such that we can bypass the bias-correction step (which is anyway irrelevant for integral validation). The optimal SNR of each source is $\rho_{\rm opt} \approx 123$, such that the LSA holds.

With the LSA likelihood~\eqref{eq:normallikelihood}, the Monte Carlo integral approximation of Eq.~\eqref{eq:sourceintegral} is~\cite{robert_monte_2004}
\begin{align}
    p(\hpsi_i|\vlambda,\mathH) \approx \frac{V}{N}\sum_{n = 1}^N \mathcal{N}(\vpsi_n|\hpsi,(\Gamma^{\psi\psi})^{-1})\pi(\vpsi_n|\vlambda,\mathH)
\end{align}
where $\{\vpsi_n\}$ is a set of $N$ random samples drawn uniformly from a suitably large hypercube of volume $V$ around the inferred parameters $\hpsi$. The corresponding approximate hyperlikelihoods for the three hypotheses were derived above. To gauge the validity of our results, we define the ratio $r^{\rm MC}_{\rm approx}$ of the Monte Carlo integral to the approximate hyperlikelihood in each hypothesis, such that $r^{\rm MC}_{\rm approx} = 1$ implies an exact match between the two approaches. As we show in Fig.~\ref{fig:integralvalidation}, $r^{\rm MC}_{\rm approx}$ is consistent with unity for large $N$ in all three hypotheses up to stochastic errors in the Monte Carlo integral, validating the constructed approximate hyperlikelihoods. We emphasize that, given the FIM $\Gamma$ at $\vtheta^*$ and the MLE $\hpsi$, the analytical hyperlikelihood evaluation is $\mathcal{O}(N)$ more efficient than the Monte Carlo approach for \textit{each} source in the population, underscoring the relevance of such approximations for the feasibility of our analysis.

\section{Description of parameters used in the results}~\label{app:resulttables}

The hyperparameters used to generate the four populations $P_v$, $P_\ell$, $P_g$, and $P_{\rm mix}$ in Sec.~\ref{sec:populationanalysis} are presented in Table~\ref{tab:truehyper}. The corresponding priors on all source parameters are summarized in Table~\ref{tab:priorsall}, and the priors on the population hyperparameters used in the inference are summarized in Table~\ref{tab:hyperpriorsall}.

\begin{table*}
    \centering
    \begin{tabular}{c c c c c}
        \hline
        \textbf{hyper-} & \multirow{2}{*}{$\boldsymbol{P_v}$} & \multirow{2}{*}{$\boldsymbol{P_\ell}$} & \multirow{2}{*}{$\boldsymbol{P_g}$} & \multirow{2}{*}{$\boldsymbol{P_{\rm mix}}$} \\
        \textbf{parameter} & & & & \\
        \hline
        \hline
        & & \\
        $K^*$ & $0.005$ & $0.005$ & $0.005$ & $0.005$ \\
        $\alpha^*$ & 0.0 & 0.0 & 0.0 & 0.0 \\
        $\beta^*$ & 0.0 & 0.0 & 0.0 & 0.0\\
        $f^*$ & 0.0 & 0.5 & 0.0 & 0.5\\
        $\mu^*_{A_\ell}$ & - & $10^{-6}$ & - & $10^{-6}$\\
        $\mu^*_{n_\ell}$ & - & 8.0 & - & 8.0\\
        $\sigma^*_{A_\ell}$ & - & $10^{-7}$ & - & $10^{-7}$\\
        $\sigma^*_{n_\ell}$ & - & 1.0 & - & 1.0\\
        $\dot{G}^* [\rm yr^{-1}]$ & 0.0 & 0.0 & $10^{-12}$ & $10^{-12}$\\ 
        & & & & \\
        \hline
    \end{tabular}
    \caption{\justifying True hyperparameters in the constructed populations $P_v$, $P_\ell$, $P_g$, and $P_{\rm mix}$. $P_v$, $P_\ell$, and $P_g$ are constructed assuming $\mathH_v$, $\mathH_\ell$, and $\mathH_g$ as the true hypothesis, respectively, while $P_{\rm mix}$ is such that the full population is affected by the global time-varying $G$ effect while a fraction $f$ of the population simultaneously experiences the local migration torque effect.}
    \label{tab:truehyper}
\end{table*}

\begin{table*}
    \centering
    \begin{tabular}{ccc}
        \hline
        \multirow{2}{*}{\textbf{parameter}} & \multirow{2}{*}{\textbf{prior pdf}} & \multirow{2}{*}{\textbf{description}} \\
        & & \\
        \hline
        \hline
        & & \\
        $\ln M,z$ & Eq.~\eqref{eq:priorvacuum} & detector-frame log-MBH mass, source redshift\\
        $\log_{10}q$ & $\mathcal{U}[-5.5,-4.5]$ & log of the small-mass-ratio $q := m_2/m_1$ \\
        $\mu$ & $Mq$ & detector-frame CO mass\\
        $a$ & $\mathcal{U}[0.5,0.99]$ & MBH spin\\
        $\theta_S$ & $\mathcal{U}[0,\pi]$ & polar sky location\\
        $\theta_K$ & $\mathcal{U}[0,\pi]$ & polar spin orientation\\
        $\phi_S$ & $\mathcal{U}[0,2\pi]$ & azimuthal sky location\\
        $\phi_K$ & $\mathcal{U}[0,2\pi]$ & azimuthal spin orientation\\
        $\Phi_0$ & $\mathcal{U}[0,2\pi]$ & initial orbital phase \\
        $T_{\rm plunge}$ & $\mathcal{U}[0.5,2]$ years & time to plunge at the start of observation\\
        $p_0$ & $p_0(T_{\rm plunge})$ (fixed) & initial semi-latus rectum\\
        $A_\ell, n_\ell$ & Eq.~\eqref{eq:priorlocal} & amplitude \& slope of the (local) migration torque \\
        $A_g$ & Eq.~\eqref{eq:priorglobal} & intrinsic strength of the (global) time-varying $G$ effect\\
        $n_g$ & 4 (fixed) & PN-order of the (global) time-varying $G$ effect\\
        & & \\
        \hline
    \end{tabular}
    \caption{\justifying Priors on the source parameters used to generate the true population. $p_0 = p_0(T_{\rm plunge})$ where a CO with semi-latus rectum $p_0(T_{\rm plunge})$ plunges into the MBH at the end of $T_{\rm plunge}$ for a given set of $\{M,\mu,a\}$.}
    \label{tab:priorsall}
\end{table*}

\begin{table*}
    \centering
    \begin{tabular}{c c c}
        \hline
        \multirow{2}{*}{\textbf{hyperparameter}} & \multirow{2}{*}{\textbf{hyperprior pdf}} & \multirow{2}{*}{\textbf{description}} \\
        & & \\
        \hline
        \hline
        & & \\
        $K$ & $\mathcal{U}[0.9K^*,1.1K^*]$ & MBH population size scaling factor\\
        $\alpha$ & $\mathcal{U}[-0.1,0.1]$ & MBH mass slope\\
        $\beta$ & $\mathcal{U}[-0.1,0.1]$ & MBH redshift slope\\
        $f$ & $\mathcal{U}[0,1]$ & fractional population with the (local) migration torques\\
        $\mu_{A_\ell}$ & $\mathcal{U}[0.9\mu^*_{A_\ell},1.1\mu^*_{A_\ell}]$ & mean of the (local) migration torque amplitudes\\
        $\mu_{n_\ell}$ & $\mathcal{U}[0.9\mu^*_{n_\ell}, 1.1\mu^*_{n_\ell}]$ & mean of the (local) migration torque slopes\\
        $\sigma_{A_\ell}$ & $\mathcal{U}[0.9\sigma^*_{A_\ell},1.1\sigma^*_{A_\ell}]$ & standard deviation of the (local) migration torque amplitudes\\
        $\sigma_{n_\ell}$ & $\mathcal{U}[0.9\sigma^*_{n_\ell},1.1\sigma^*_{n_\ell}]$ & standard deviation of the (local) migration torque slopes\\
        $\dot{G}$ &  $\mathcal{U}[-5,5]\times 10^{-12}$& intrinsic strength of the (global) time-varying $G$ effect (per year)\\ 
        & & \\
        \hline
    \end{tabular}
    \caption{\justifying Hyperpriors on the population hyperparameters. $\cdot^*$ denotes the true value of the hyperparameter in the chosen population (Table~\ref{tab:truehyper}).}
    \label{tab:hyperpriorsall}
\end{table*}

\end{document}